\documentclass[prl, twocolumn]{revtex4}
\usepackage[ansinew]{inputenc}
\usepackage{graphicx}
\usepackage{amsmath}
\usepackage{amsthm}
\usepackage{bm}
\usepackage{layout}
\usepackage{float}
\usepackage{amsfonts}
\usepackage{amssymb}
\usepackage{hyperref}
\usepackage{txfonts}%
\newcommand\id{\leavevmode\hbox{\small1\kern-3.3pt\normalsize1}}

\newcommand{\tr}{\mbox{Tr}}

\begin{document}

\title{Operational formulation of time reversal in quantum theory\footnote{The published version of this paper can be found in \textit{Nature Physics} 11, 853–858 (2015), doi:10.1038/nphys3414, at \href{http://www.nature.com/nphys/journal/v11/n10/full/nphys3414.html}{http://www.nature.com/nphys/journal/v11/n10/full/nphys3414.html} .}}


\author{Ognyan Oreshkov and Nicolas J. Cerf}

\affiliation{QuIC, Ecole Polytechnique de Bruxelles, CP 165, Universit\'{e} Libre de Bruxelles, 1050 Brussels, Belgium.}

\begin{abstract}

The symmetry of quantum theory under time reversal has long been a subject of controversy because the transition probabilities given by Born's rule do not apply backward in time. Here, we resolve this problem within a rigorous operational probabilistic framework. We argue that reconciling time reversal with the probabilistic rules of the theory requires a notion of operation that permits realizations via both pre- and post-selection. We develop the generalized formulation of quantum theory that stems from this approach and give a precise definition of time-reversal symmetry, emphasizing a previously overlooked distinction between states and effects. We prove an analogue of Wigner's theorem, which characterizes all allowed symmetry transformations in this operationally time-symmetric quantum theory. Remarkably, we find larger classes of symmetry transformations than those assumed before. This suggests a possible direction for search of extensions of known physics.

\end{abstract}

\maketitle


Symmetries play a fundamental role in our understanding of physics. It is widely believed that the most general symmetry transformations in quantum theory correspond to unitary or anti-unitary transformations on the Hilbert space, with symmetries involving time reversal being anti-unitary \cite{Wigner}. This has profound implications for many phenomena, such as the classification of possible elementary particles \cite{Wigner2}. The joint transformation of charge conjugation, parity inversion, and time reversal defined according to this principle, is considered an exact symmetry of all known physical laws \cite{Schwinger, Lueders, Pauli, BellCPT}. However, it has been recognized that Born's rule, which describes the probabilities for the outcomes of future measurements conditional on past preparations, does not apply for events in the reverse order \cite{Watanabe, ABL}. This is in conflict with the very definition of symmetry underlying the above assertions \cite{Holster}. Moreover, since the operational interpretation of a quantum state is directly linked to Born's rule \cite{Busch}, this raises doubts about whether the commonly accepted notion of time-reversed state makes physical sense.

Here, we address this problem from a rigorous operational perspective \cite{Ludwig, Hardy, Barrett, HardyCircuit,CDP2, DB, MM, CDP, Hardy2, infounit}, using the circuit framework for operational probabilistic theories (OPTs) \cite{HardyCircuit,CDP2}, which has been shown to successfully formalize the informational foundations of quantum theory \cite{CDP, Hardy2}. We argue that reconciling time reversal with the probabilistic rules of the theory requires a generalized notion of operation, defined without assumptions on whether the implementation of an operation involves pre- or post-selection. In this approach, operations are not expected to be up to the `free choices' of agents, but merely describe knowledge about the possible events taking place in different regions, conditional on information obtained locally. We develop the generalized formulation of quantum theory that stems from this approach and show that it has a new notion of state space that is not convex. We give a precise definition of time reversal symmetry, taking into account the different nature of states and effects, which has been overlooked in previous treatments. We prove an analogue of Wigner's theorem \cite{Wigner}, which characterizes all possible symmetry transformations in this time-symmetric formulation of quantum theory. Remarkably, we find more general classes of symmetry transformations than those assumed before. 

We also identify the time asymmetry ingrained in the standard formulation of quantum theory as the fact that, forward in time, without post-selection we can only prepare a restricted class of all allowed operations, which does not hold backward in time. We show how this property can be expressed formally in the circuit framework, and that it can be understood as a result of the unitarity of the dynamics in space-time combined with the form of the past and future boundary operations. We establish an exact link between this asymmetry and the fact that we can remember the past but not the future. 

\section*{The circuit framework}

The basic concept in the circuit framework for OPTs \cite{HardyCircuit, CDP2, CDP, Hardy2} is that of \textit{operation}, corresponding to `one use of a physical device with an input and an output system'. An operation with an input system $A$ and an output system $B$ is described by a collection of events $\{\mathcal{M}^{A\rightarrow B}_i\}_{i\in O}$ labeled by an outcome index $i$ taking values in some set $O$. Pictorially, operations are represented by `boxes' with input and output `wires' (Fig. 1). Operations whose input system is trivial (depicted with no wire) are called \textit{preparations}, and those whose output system is trivial are called \textit{measurements} (the trivial system is denoted by $I$). Operations can be composed in sequence and in parallel yielding new operations \cite{Coecke} (see Supplementary Methods). A \textit{circuit} is an acyclic composition of operations with no open wires  (Fig. 2). The central idea of the circuit framework is that a theory prescribes joint probabilities for the operation outcomes in every given circuit, which depend only on the specification of the circuit \cite{HardyCircuit,Hardy2}. Equivalently \cite{CDP2, CDP}, for any preparation $\{\rho^{I\rightarrow A}_i\}_{i\in O}$ and any measurement $\{E^{A\rightarrow I}_j \}_{j\in Q}$, the theory prescribes joint probabilities $p(i,j|\{\rho^{I\rightarrow A}_i\}_{i\in O}, \{E^{A\rightarrow I}_j \}_{j\in Q})\geq 0$, $\sum_{i\in O,j\in Q}p(i,j|\{\rho^{I\rightarrow A}_i\}_{i\in O}, \{E^{A\rightarrow I}_j \}_{j\in Q})=1$, where for parallel circuits the probabilities factor out.

\renewcommand{\figurename}{{Fig.}}

\begin{figure}
\vspace{0.5 cm}
\begin{center}
\includegraphics[width=3.5cm]{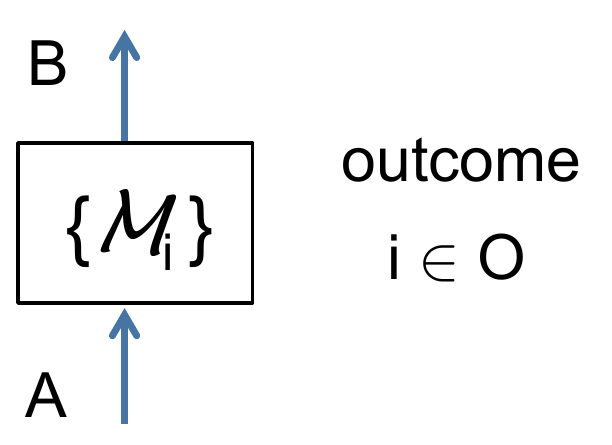}
\end{center}
\vspace{-0.5 cm}
\caption{\textbf{Operation.} In the circuit framework for operational probabilistic theories, an operation $\{\mathcal{M}^{A\rightarrow B}_i\}_{i\in O}$ is defined as a collection of possible events from an input system $A$ to an output system $B$, labeled by an outcome index ${i\in O}$. In the standard approach, an operations is implicitly assumed to be realized without post-selection, while our generalized formulation permits both pre- and post-selection. } \label{operation}
\end{figure}

A circuit formalizes the idea of information exchange mediated by systems \cite{Coecke}. By definition, the wires in a circuit are the only means of information exchange responsible for the correlations between the events in the boxes. One can figuratively think of the boxes in a circuit as isolated space-time regions, which can communicate with each other only through the wires (see Supplementary Methods). The description of the operation in a given box is determined only based on variables in that box. 

An OPT is completely defined by specifying all possible operations and the probabilities for the outcomes of all possible circuits. It is formulated in terms of equivalence classes of operations---if two operations $\{\mathcal{M}^{A\rightarrow B}_i\}_{i\in O}$ and $\{\mathcal{N}^{A\rightarrow B}_i\}_{i\in O}$ yield the same joint probabilities in all circuits that they may be plugged in, they are deemed equivalent. Similarly, one defines equivalence classes of events $\mathcal{M}^{A\rightarrow B}_i\in\{\mathcal{M}^{A\rightarrow B}_i\}_{i\in O}$ and $\mathcal{N}^{A\rightarrow B}_j\in\{\mathcal{N}^{A\rightarrow B}_j\}_{j\in Q}$  that may belong to different operations. They are called \textit{transformations} \cite{CDP2}. In the case of preparation and measurement events, they are called \textit{states} and \textit{effects}, respectively \cite{Ludwig}. The joint probabilities of preparation and measurement events are then functions of the state and effect only, $p(i,j|\{\rho^{I\rightarrow A}_i\}_{i\in O}, \{E^{A\rightarrow I}_j\}_{j\in Q}) = p(\rho^{I\rightarrow A}_i, E^{A\rightarrow I}_j)$. States are thus real functions on effects and vice versa.

\begin{figure}
\vspace{0.5 cm}
\begin{center}
\includegraphics[width=8.5cm]{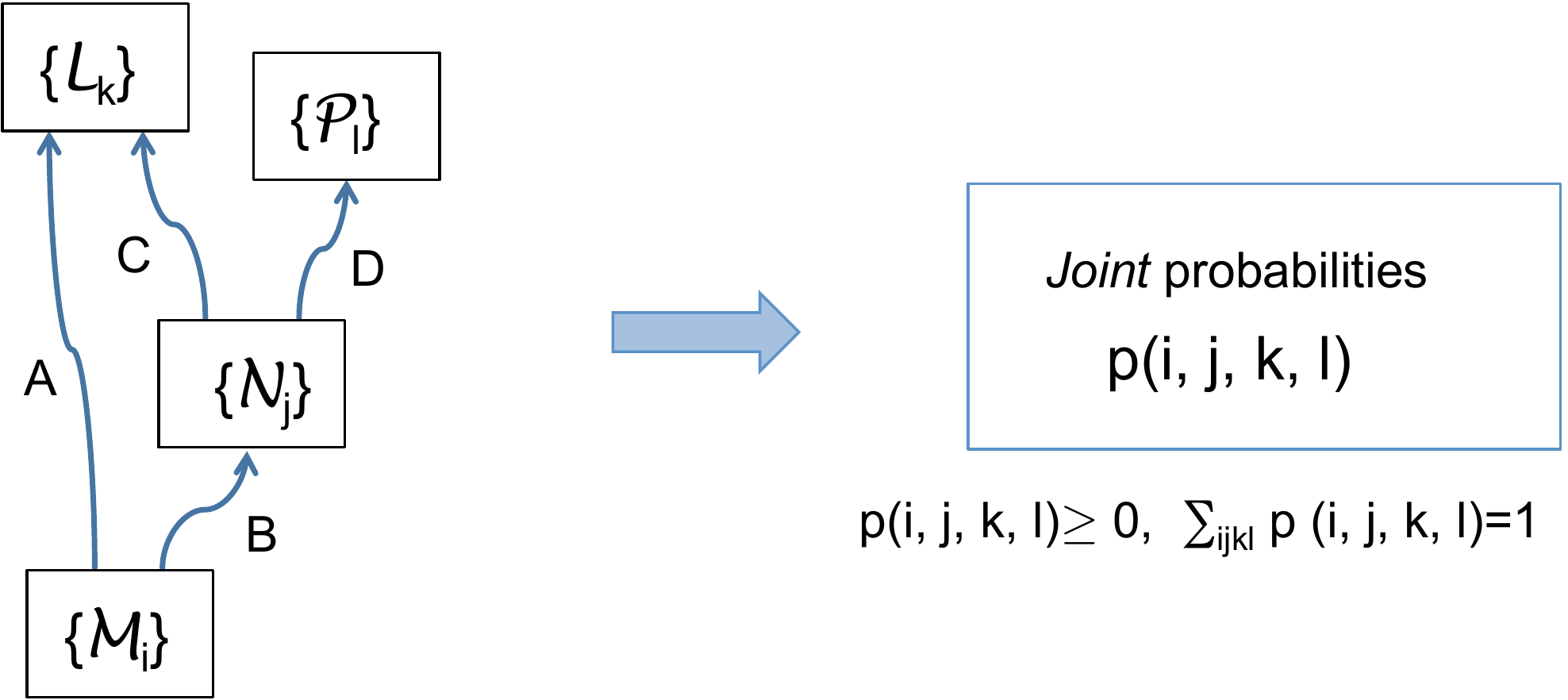}
\end{center}
\vspace{-0.5 cm}
\caption{\textbf{Circuit.} A circuit is an acyclic composition of operations with no open wires. An operational probabilistic theory prescribes joint probabilities for the outcomes of any given circuit \cite{HardyCircuit,CDP2,CDP,Hardy2}.} \label{circuit}
\end{figure}

Quantum theory is a special case of OPT, in which a system $A$ is associated with a Hilbert space $\mathcal{H}^A$ of dimension $d_A$ (we assume finite dimensions), and a transformation from $A$ to $B$ is a completely positive (CP) and trace-nonincreasing linear map $\mathcal{M}^{A\rightarrow B}: \mathcal{L}({\mathcal{H}^A})\rightarrow \mathcal{L}({\mathcal{H}^B})$, where $\mathcal{L}({\mathcal{H}^X})$ denotes the space of linear operators over $\mathcal{H}^X$. (The Hilbert space of a composite system $XY$ is the tensor product $\mathcal{H}^X\otimes \mathcal{H}^Y$). A standard quantum operation is a collection of CP maps $\{\mathcal{M}^{A\rightarrow B}_i\}_{i\in O}$, whose sum $\sum_{i\in O} \mathcal{M}^{A\rightarrow B}_i = \overline{\mathcal{M}}^{A\rightarrow B}$ is a CP and trace-preserving map. Using a convenient isomorphism, states $\rho^{I\rightarrow A}$ and effects $E^{A\rightarrow I}$ are represented as positive semidefinite (PS) operators $\rho^A, E^A, \in \mathcal{L}(\mathcal{H})$ (see Supplementary Methods). In particular, a preparation is described by a set of PS operators $\{\rho^A_i\}_{i\in O}$, such that $\sum_{i\in O}\tr(\rho^A_i) = 1$, and a measurement by a set of PS operators $\{E^A_j\}_{j\in Q}$, such that $\sum_{j\in Q} E^A_j=\id^A$. The joint probabilities of states and effects are then given by
\begin{gather}
p(\rho^{I\rightarrow A}_i, E^{A\rightarrow I}_j) 
= \tr(\rho^A_i E^A_j). \label{standardrule}
\end{gather}

\section*{Causality and the no-post-selection criterion}

Operations in the standard formulation of quantum theory obey the axiom of \textit{causality} \cite{CDP2,CDP}, which says that for every preparation $\{\rho^{I\rightarrow A}_i\}_{i\in O}$ composed with a measurement $\{E^{A\rightarrow I}_j\}_{j\in Q}$, the probabilities of the preparation events do not depend on the measurement, i.e., $\sum_{j\in O}p(i,j|\{\rho^{I\rightarrow A}_i\}_{i\in O}, \{E^{A\rightarrow I}_j\}_{j\in Q}) \equiv p(i|\{\rho^{I\rightarrow A}_i\}_{i\in O})$ is the same for every $\{E^{A\rightarrow I}_j\}_{j\in Q}$. This implies that the outcomes of past operations in a circuit do not depend on operations in the future. But the outcomes of future operations can depend on past operations, which shows an explicit time asymmetry in the standard formulation of quantum theory. 

The essence of this asymmetry can be understood by observing that an operation is implicitly assumed to be realized \textit{without post-selection}, i.e., the occurrence of an operation is assumed conditional only on information available prior to the time of the input system (see Supplementary Methods). Indeed, conditionally on information available in the future, we could obtain non-standard `operations' in a given box, which violate the axiom. 
Thus, the axiom expresses a non-trivial constraint on the operations obtainable by pre-selection only.

An accurate comparison between the forward and backward directions of time requires identifying the `pre-selected' operations in the backward direction. They correspond to the sets of possible events in a given box that can be known to have occurred in the past in the forward direction, and include all subsets of the outcomes of standard operations. Thus, there is a physical asymmetry concerning the fact that the operations that can be obtained without post-selection in the two directions of time are different. The origin of this asymmetry will be analyzed later.

A time-symmetric theory should describe observations in both directions of time via the same rules. Since the events that correspond to valid operations according to the no-post-selection criterion respect the causality axiom in one direction but not the other, based on this criterion there does not exist an empirically confirmed time-symmetric theory that agrees with the circuit connections assumed in the standard theory. Without \textit{ad hoc} assumptions, the only way of obtaining such a theory is to drop the no-post-selection criterion from the definition of operation. 

We therefore propose to view an operation simply as a description of the possible events in a given box, conditional on information obtained without looking into other boxes in the circuit, irrespectively of when in time this information is available. (We will show that any constraints on the latter follow from the form of the dynamics in space-time.) From this perspective, learning or discarding of information about the outcomes of an operation should yield another valid operation in agreement with the corresponding update of the probabilities in a circuit. Intuitively, one may imagine that the information about the events in each box in a circuit is stored in a separate `safe' that can be opened in the future \cite{Molmer}. Upon looking into the content of a given safe, an experimenter will update her description of its content, as well as the probabilities for the contents of other safes. We next present the generalized formulation of quantum theory that follows from this point of view.

\section*{Generalized formulation}

The generalized formulation is summarized by the following rules (see derivation in Methods).

Equivalent operations are described by a collection of CP maps $\{\mathcal{M}^{A\rightarrow B}_j \}_{j\in O}$, where $\overline{\mathcal{M}}^{A\rightarrow B}= \sum_{i\in O}\mathcal{M}^{A\rightarrow B}_i$ satisfies
\begin{gather}
\textrm{Tr}(\overline{\mathcal{M}}^{A\rightarrow B}(\frac{\id^A}{d_A}))=1. \label{Mbar}
\end{gather}
(Note that $\overline{\mathcal{M}}^{A\rightarrow B}$ does not have to be trace-preserving.)

The sequential composition of two operations $\{\mathcal{M}^{A\rightarrow B}_i\}_{i\in O}$ and  $\{\mathcal{N}^{B\rightarrow C}_j\}_{j\in Q}$ is a new operation $\{\mathcal{L}^{A\rightarrow C}_{ij}\}_{i\in O,j\in Q}$, where
 \begin{gather}
 \mathcal{L}^{A\rightarrow C}_{ji} =  \frac{ \mathcal{N}^{B\rightarrow C}_{j} \circ\mathcal{M}^{A\rightarrow B}_{i}} { \tr (\overline{\mathcal{N}}^{B\rightarrow C}\circ \overline{\mathcal{M}}^{A\rightarrow B}(\frac{\id^A}{d_A}))}, \hspace{0.2cm} i\in O, j \in Q,\label{seqcomp}
 \end{gather}
unless $\overline{\mathcal{N}}^{B\rightarrow C} \circ \overline{\mathcal{M}}^{A\rightarrow B} = {0}^{A\rightarrow C}$, where ${0}^{A\rightarrow C}$ is the null CP map. In the latter case, the composition never occurs, or, equivalently, its result is the \textit{null} operation from $A$ to $C$.

As in the standard formulation \cite{CDP2,CDP}, CP maps from the trivial system to itself are interpreted as probabilities. Since every circuit is equivalent to an operation from the trivial system to itself, the above rules define the probabilities for all circuits.

Importantly, the equivalent events, or \textit{transformations}, are not given by the CP maps above. They are described by \textit{pairs} of CP maps, $(\mathcal{M}^{A\rightarrow B}; \overline{\mathcal{M}}^{A\rightarrow B})$, with the property
\begin{gather}
\mathcal{M}^{A\rightarrow B}(\rho^A) \leq \overline{\mathcal{M}}^{A\rightarrow B}(\rho^A), \forall \rho^A\geq 0, \hspace{0.05cm}\textrm{Tr}(\overline{\mathcal{M}}^{A\rightarrow B}(\frac{\id^A}{d_A}))=1. \label{transformations}
\end{gather}

\textit{States} are represented by $(\rho^A;\overline{\rho}^A)$, $\rho^A\leq \overline{\rho}^A$, $\tr(\overline{\rho}^A)=1$, $\rho^A, \overline{\rho}^A \in \mathcal{L}(\mathcal{H}^A)$, and \textit{effects} by $(E^A;\overline{E}^A)$, $E^A\leq \overline{E}^A$, $\tr(\overline{E}^A)=d_A$, $E^A, \overline{E}^A \in \mathcal{L}(\mathcal{H}^A)$, with the main probability rule reading
\begin{gather}
p\left((\rho^A;\overline{\rho}^A), (E^A;\overline{E}^A)\right ) =
\frac{\tr ( \rho^A E^A)} { \tr (\overline{\rho}^A  \overline{E}^A)},\hspace{0.2cm} \textrm{for} \hspace{0.1cm}  \tr (\overline{\rho}^A  \overline{E}^A)\neq 0,\notag\\
\hspace{1.9cm} =0,\hspace{0.2cm} \textrm{for} \hspace{0.1cm}  \tr (\overline{\rho}^A  \overline{E}^A)= 0.
\label{state-effect}
\end{gather}
(Born's rule is obtained for $\rho=\overline{\rho}$, $\overline{E}=\id$.)

Notably, the sets of states and effects, viewed as real functions on each other via Eq.~\eqref{state-effect}, are not closed under convex combinations. The convex combinations of these functions do not correspond to events that can be obtained by local procedures in the preparation and measurement boxes (see Methods).

The most general rule for updating an operation upon learning or discarding of information is presented in Methods. 

We remark that the approach we have proposed is not limited to quantum theory. In particular, it can be used to generalize any OPT formulated in the standard approach. In the Supplementary Methods, we illustrate the case of classical OPT with an example. 


\section*{Time reversal and general symmetries}

Under time reversal $\mathcal{T}$, every operation $\{\mathcal{M}^{A\rightarrow B}_i\}_{i\in O}$ is expected to be seen as an operation $\{\tilde{\mathcal{M}}^{B\rightarrow A}_i\}_{i\in O}$, such that the probabilities of any circuit under this map $\mathcal{T}$ remain invariant. In particular, states should become effects and vice versa, such that their joint probabilities are preserved. There are, however, infinitely many transformations with this property. Time reversal is a specific map between the spaces of states and effects, which is determined by the laws of mechanics and should be understood in the following sense. Imagine that we could create a measurement box whose classical description looks just like that of a given preparation box operating in reverse temporal order. The measurement implemented by the measurement box is the time-reversed image of the preparation implemented by the preparation box (Fig. 3). Before we discuss how the two can be related, we give a characterization of all possible symmetry transformations, i.e., all transformations of boxes that leave the probabilities of circuits invariant.

\begin{figure}
\vspace{0.5 cm}
\begin{center}
\includegraphics[width=8cm]{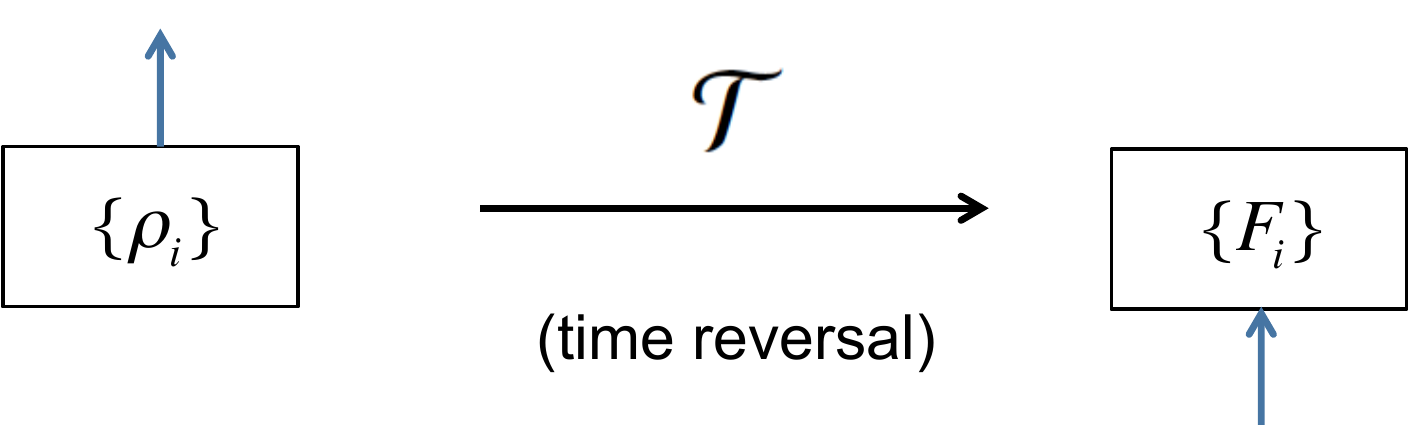}
\end{center}
\vspace{-0.5 cm}
\caption{\textbf{Time reversal as an active transformation.} If we could actively `flip' the time orientation of a preparation box (e.g., create a process that looks just like the preparation process played in reverse), we would obtain a measurement box. The measurement implemented by that box, characterized relative to preparations that have not been `flipped', is the time-reversed image of the preparation.} \label{TimeFlip}
\end{figure}

A crucial insight in our analysis is that states and effects on the same systems $A$ are \textit{distinct} objects---they are associated with separate events and live in separate spaces, $\textrm{St}^A$ and $\textrm{Eff}^A$, even though we describe them using operators in the same space $\mathcal{L}(\mathcal{H}^A)$. Importantly, the latter is based on a canonical isomorphism which merely reflects a choice of representation of the pairing between dual vectors, $(\rho^{I\rightarrow A}, E^{A\rightarrow I}) = \tr[\rho^AE^A]$ (see Methods). Therefore, a symmetry transformation $\mathcal{S}^A$ must be defined by its action on both spaces, $\mathcal{S}^A:\textrm{St}^A\times \textrm{Eff}^A\rightarrow \textrm{St}^A\times \textrm{Eff}^A$. We can distinguish two types of symmetry transformations. \textit{Type} $I$---those that map states to states and effects to effects (e.g., spatial rotation). They can be thought of as consisting of a pair of transformations,   $\mathcal{S}^A_{s\rightarrow s}: \textrm{St}^A\rightarrow \textrm{St}^A$ and $\mathcal{S}^A_{e\rightarrow e}: \textrm{Eff}^A\rightarrow \textrm{Eff}^A$, whose form in the canonical representation will be denoted by  $\hat{S}^A_{s\rightarrow s}$, $\hat{S}^A_{e\rightarrow e}$. \textit{Type} $II$---those that map states to effects and effects to states (e.g., time reversal). They can be thought of as consisting of a pair of transformations,   $\mathcal{S}^A_{s\rightarrow e}: \textrm{St}^A\rightarrow \textrm{Eff}^A$ and $\mathcal{S}^A_{e\rightarrow s}: \textrm{Eff}^A\rightarrow \textrm{St}^A$, whose canonical representation will be denoted by $\hat{S}^A_{s\rightarrow e}$, $\hat{S}^A_{e\rightarrow s}$. Hereafter, we drop the superscript $A$.

\textit{Theorem}. Consider a system with Hilbert space $\mathcal{H}$ of dimension $d$. Symmetry transformations of {\textit{type} $I$} have the form
\begin{gather}
\hat{S}_{s\rightarrow s}( \rho; \overline{\rho}) = (\sigma; \overline{\sigma}) = (\frac{S\rho S^{\dagger}}{\tr(S\overline{\rho} S^{\dagger})};  \frac{S\overline{\rho} S^{\dagger}}{\tr(S\overline{\rho} S^{\dagger})}     ), \label{th1}\\
\hat{S}_{e\rightarrow e}( E; \overline{E}) = (F; \overline{F}) = (d\frac{{S^{-1}}^{\dagger}E S^{-1}}{\tr({S^{-1}}^{\dagger}\overline{E} S^{-1})}; d\frac{{S^{-1}}^{\dagger}\overline{E} S^{-1}}{\tr({S^{-1}}^{\dagger}\overline{E} S^{-1})}  ),\label{th2}
\end{gather}
or the form
\begin{gather}
\hat{S}_{s\rightarrow s}( \rho; \overline{\rho}) = (\sigma; \overline{\sigma}) = (\frac{S\rho^T S^{\dagger}}{\tr(S\overline{\rho}^T S^{\dagger})};  \frac{S\overline{\rho}^T S^{\dagger}}{\tr(S\overline{\rho}^T S^{\dagger})}     ), \label{th3}\\
\hat{S}_{e\rightarrow e}( E; \overline{E}) = (F; \overline{F})= (d\frac{{S^{-1}}^{\dagger}E^T S^{-1}}{\tr({S^{-1}}^{\dagger}\overline{E}^T S^{-1})}; d\frac{{S^{-1}}^{\dagger}\overline{E}^T S^{-1}}{\tr({S^{-1}}^{\dagger}\overline{E}^T S^{-1})}  ),\label{th4}
\end{gather}
where $T$ denotes transposition in some basis, and $S\in \mathcal{L}(\mathcal{H})$ is an invertible operator. Similarly, symmetry transformations of {\textit{type} $II$} have the form
\begin{gather}
\hat{S}_{s\rightarrow e}( \rho; \overline{\rho}) = (F; \overline{F})= (d\frac{S\rho S^{\dagger}}{\tr(S\overline{\rho} S^{\dagger})}; d \frac{S\overline{\rho} S^{\dagger}}{\tr(S\overline{\rho} S^{\dagger})}     ),\label{Th1} \\
\hat{S}_{e\rightarrow s}( E; \overline{E})  = (\sigma; \overline{\sigma}) =(\frac{{S^{-1}}^{\dagger}E S^{-1}}{\tr({S^{-1}}^{\dagger}\overline{E} S^{-1})}; \frac{{S^{-1}}^{\dagger}\overline{E} S^{-1}}{\tr({S^{-1}}^{\dagger}\overline{E} S^{-1})}  ),\label{Th2}
\end{gather}
or the form
\begin{gather}
\hat{S}_{s\rightarrow e}( \rho; \overline{\rho}) = (F; \overline{F})= (d\frac{S\rho^T S^{\dagger}}{\tr(S\overline{\rho}^T S^{\dagger})}; d \frac{S\overline{\rho}^T S^{\dagger}}{\tr(S\overline{\rho}^T S^{\dagger})}     ), \label{Th3} \\
\hat{S}_{e\rightarrow s}( E; \overline{E}) = (\sigma; \overline{\sigma}) = (\frac{{S^{-1}}^{\dagger}E^T S^{-1}}{\tr({S^{-1}}^{\dagger}\overline{E}^T S^{-1})}; \frac{{S^{-1}}^{\dagger}\overline{E}^T S^{-1}}{\tr({S^{-1}}^{\dagger}\overline{E}^T S^{-1})}  ). \label{Th4}
\end{gather}
This implies the transformation of arbitrary operations. The proof is presented in Methods.

If we assume that an isolated system must follow unitary evolution driven by a Hamiltonian, and energy should not change under time reversal, we obtain that time reversal must be described by a transformation of the form \eqref{Th3} and \eqref{Th4} (see Methods). If this is to hold for arbitrary Hamiltonians, then $S$ must be unitary. The concrete $S$, which depends on the transposition basis, would be determined by how specific observables transform under time reversal (e.g., energy remains invariant, spin changes sign). Note that since the generalized formulation permits more general than unitary evolutions, transformations with non-unitary $S$ are also conceivable in principle.

The original classification of symmetries by Wigner \cite{Wigner} is obtained within the traditional exposition of quantum theory, where one does not distinguish states and effects but speaks of transition probabilities between states only. If we were to similarly interpret the canonical representations of effects in the state space (times $\frac{1}{d}$) as states, a symmetry transformation would be described by a single map from states to states and we would obtain that $S$ must be unitary. The operators $\rho\in \mathcal{L}(\mathcal{H})$ would then transform as either $\rho \rightarrow S\rho S^{\dagger}$ or $\rho \rightarrow S\rho^T S^{\dagger}$, which amounts respectively to a unitary or an anti-unitary transformation on the vectors in the underlying Hilbert space, which agrees with Wigner's theorem. However, from an operational perspective there is no justification for this identification of states with effects. 
Furthermore, the traditional conclusion about the form of time reversal \cite{Wigner} is derived assuming that the transition probabilities given by Born's rule remain invariant under time reversal. In practice, such a transition corresponds to a measurement event following a preparation event, and the conditional probabilities of events in the reverse order are not generally described by Born's rule. This is why, in our view, the traditional conclusion is not justified. In contrast, the generalized formulation developed here gives an empirically consistent definition of time-reversal symmetry. However, it also shows the possibility for transformations with non-unitary $S$.

\section*{Understanding the observed asymmetry}

\begin{figure}
\vspace{0.5 cm}
\begin{center}
\includegraphics[width=8.0cm]{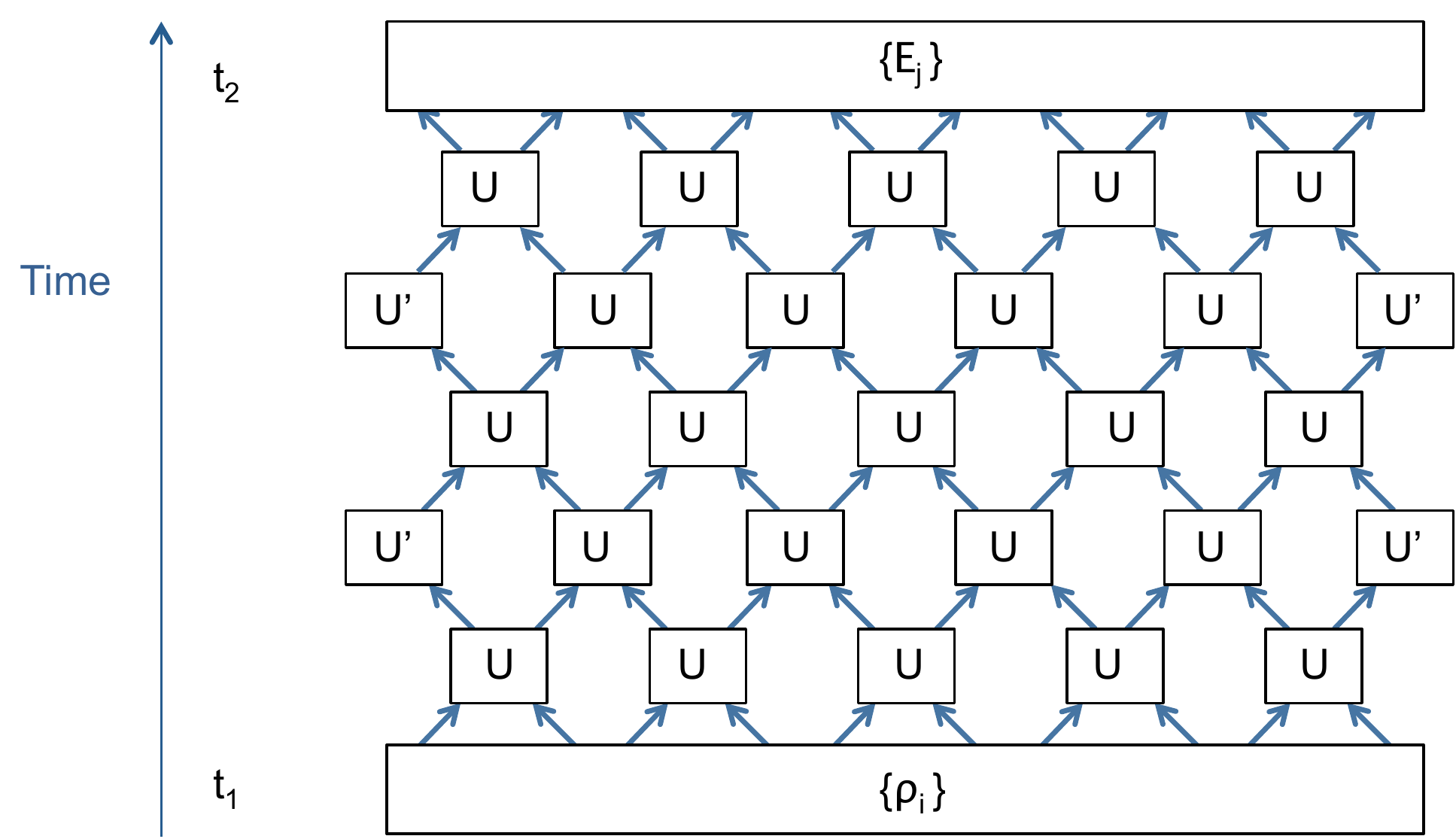}
\end{center}
\vspace{-0.5 cm}
\caption{\textbf{A toy model of the universe between two instants of time.} According to the known laws of quantum mechanics, physical systems undergo unitary evolution in time. All information about the events in the universe between times $t_1$ and $t_2$ is then encoded in the outcomes of operations on the boundaries of this space-time region. The information available at time $t_1$ is contained in the preparation box. An observer at $t_1$ can have direct access only to this information but not to the information in the measurement box. According to such an observer, all future circuits consist of standard operations if and only if the final measurement $\{E_j\}_{j\in Q}$ satisfies $\sum_{j\in Q}E_j =\id$.} \label{asymmetry1}
\end{figure}

We now investigate why without post-selection in the forward direction of time we can only implement standard quantum operations, which does not hold backward in time.

Assume that isolated systems evolve unitarily forward in time, as prescribed by the known laws of quantum mechanics. This means that if we consider all systems in the universe between times $t_1$ and $t_2$, $t_1<t_2$, we can describe their evolution by a unitary circuit (or one joint unitary operation), such that the classical information about all events between the two times is encoded in the outcomes of operations on the past and future boundaries of the circuit (Fig. 4). By definition, all information available prior to $t_1$ is contained in the preparation box (the box can be imagined to extend to the infinite past). An observer inside that box can update the description of the events in the box, but not of the events in future boxes. So according to an observer prior to $t_1$, the future events in the universe would look as in Fig. 4, where the preparation may be updated, but the final measurement is fixed. Any effective circuit in some region in the future between times $t_1$ and $t_2$ according to this observer must be consistent with the big unitary circuit, i.e., all future circuits should be possible to purify, by including the devices and environments in the description, to the circuit in Fig. 4. It is well known that if the effective circuits consist of standard operations, their unitary purification can be done with a final measurement that is a standard quantum measurement. Reversely, if every future circuit must consist of standard operations, the final measurement in particular must be a standard measurement. In other words, the claim that all future circuits that can be known at a given time unconditionally on future events must be standard quantum circuits is equivalent to a statement about the form of the future boundary operation in the circuit of the universe. This boundary operation can be moved arbitrarily far into the future, transforming it consistently with the unitary dynamics.

To analyze the time-reversed situation, assume for simplicity that time reversal is described by Eqs.~\eqref{Th3} and \eqref{Th4} with $S=\id$ (the exact form of time reversal does not affect the probabilities of events). In this case, the reverse evolution is unitary and we have a similar picture to the previous case, but with a possibly non-standard future measurement. As argued earlier, the causality axiom does not hold for pre-selected operations in the reverse direction, which means that the `future' measurement backward in time indeed cannot be a standard one. This means that the preparation in the past boundary of the universe in the forward direction (Fig. 4) cannot give the maximally mixed state on average.

If the preparation gave the maximally mixed state on average, we could not have memory of the past consistently with unitarity, because at any time, systems coming from the past would be uncorrelated. By the same token, the form of the future boundary measurement implies that we cannot have memory of the future. This establishes an exact link between the psychological \cite{Hawking, Brun} and `quantum' arrows of time.

Interestingly, Fig. 4 contains the possibility that not all classical information available in the past is available in the future, just like there is information in the future that cannot be known in the past. One can think of this as information that is lost in the forward direction of time.

By considering the purification of circuits on larger systems, we can understand the mechanism by which information about these circuits reaches different space-time locations. For example, Fig. 5 illustrates how a non-standard future boundary measurement can lead to present information about non-standard local operations in the future. In the Supplementary Methods, we discuss the consistency of the interpretation of the theory in these more general cases. 

\begin{figure}
\vspace{0.5 cm}
\begin{center}
\includegraphics[width=8cm]{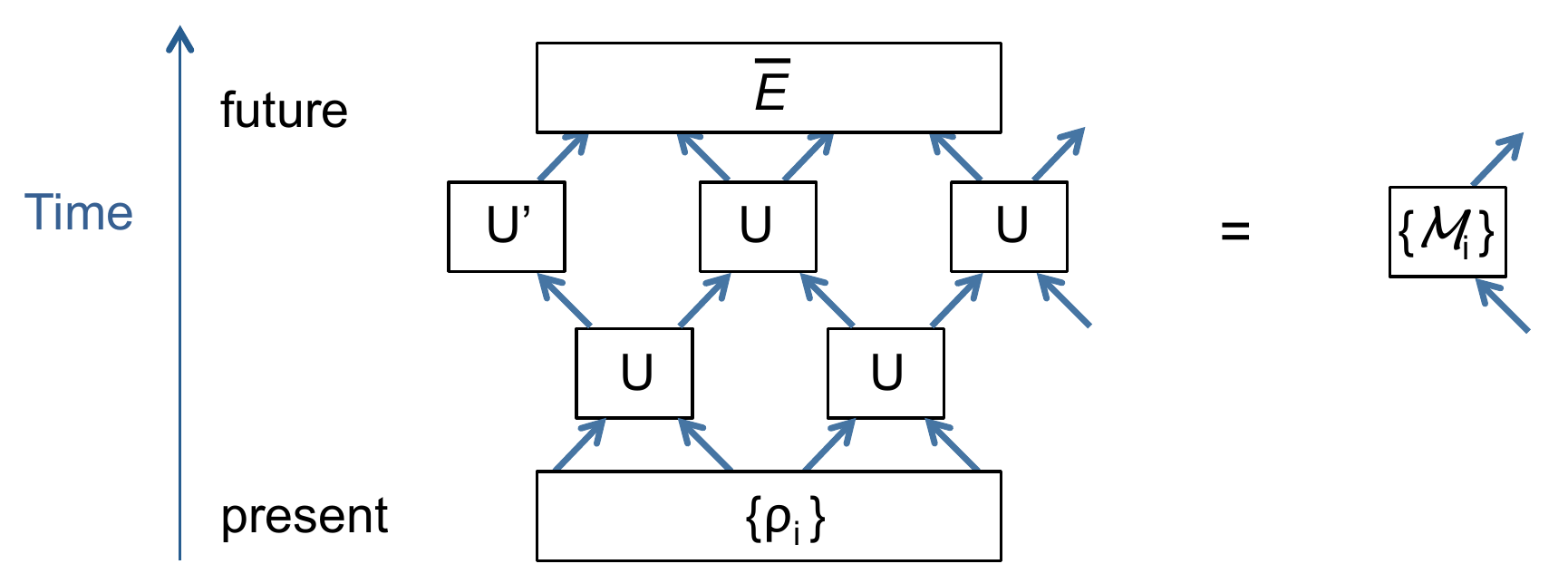}
\end{center}
\vspace{-0.5 cm}
\caption{\textbf{Preparing non-standard operations without post-selection in a world with non-standard future boundary condition.} A future boundary measurement with operator $\overline{E}\neq \id$ can allow present knowledge of future CP maps $\{\mathcal{M}_i\}_{i\in O}$ that are not proportional to CPTP maps.} \label{divination}
\end{figure}

Unlike previous models of quantum mechanics with past and future boundary conditions \cite{AR, Griffiths, Gell-MannHartle}, our approach does not interpret the future condition as a constraint on the future state of the universe, but on the future effects. It gives an explicit picture of the flow of information in space-time, where classical information by definition lives on the boundary. 

\section*{Discussion}

We have argued that an empirically consistent notion of time-reversal symmetry in quantum theory requires a generalized notion of operation, whose implementation can involve both pre- and post-selection. This has allowed us to give a rigorous definition of time-reversal symmetry based on the preservation of probabilities of events. The operational approach provides a different understanding of the accepted notion of time reversal: it is a map between two separate spaces---those of states and effects---and not from the space of states to itself. This has revealed the possibility for symmetry transformations beyond the standard classes of unitary and anti-unitary transformations predicted by Wigner's theorem. Could such symmetries be realized in nature?

One possibility is that they may arise in a novel sense in scenarios defined through post-selection, still in agreement with the known laws of quantum mechanics. Another possibility is that they may be relevant in new physical regimes, such as those where both quantum theory and gravity play a role. Indeed, these symmetry transformations, like the most general evolutions permitted in the time-symmetric formulation, are post-selection-like transformations of the kind proposed to model the dynamics of quantum systems in the presence of black holes and closed time-like curves \cite{Hartle, HorMald, GotPre, Bennett2, Svetlichny, Lloyd}. Such models are often referred to as `non-linear' extensions of quantum theory, but we have seen that this is not precise, because the notion of state in the extended theory is not the same as the standard one, and the state space is not convex. We believe that this insight is an important stepping stone for the understanding of such models.

The time-symmetric approach to the notion of operation proposed here is also conceptually suited for theories with no background time, as in the context of gravity. Building upon recent ideas for quantum theory with indefinite causal structure \cite{hardyqg, chiribella3, OCB}, the present formulation can be extended to an operational quantum theory without any predefined time \cite{OC2014}. Our demonstration that the circuit notion of causality can be regarded as non-fundamental offers a new perspective on the role of causal structure in quantum mechanics \cite{hardyqg, chiribella3, OCB, OC2014, LeiferSpekkens, Arrighi, ColbeckRenner, CoeckeLal, WoodSpekkens, Fritz2, Bancal,  Rudolph, Baumeler2, brukner, Coecke3, Pienaar, Ried}.



\section{Methods}

\subsection{Derivation of the generalized formulation}

We consider as a valid operation any set of possible events that can be obtained by a local procedure inside a box with an input and an output system, without assumptions on whether the procedure involves pre- or post-selection. If a set of events defines a valid operation in this sense, so does any subset of this set, because any subset can be selected inside the box. We assume that standard quantum theory holds, and we derive its generalized formulation based on this principle without additional assumptions. 

Consider a preparation box implementing the standard preparation $\{\rho^A_i\}_{i\in O}$, which is connected via the system $A$ to a measurement box implementing the standard measurement $\{E^A_j\}_{j\in Q}$ (we use the representation of preparations and measurements in terms of PS operators in $ \mathcal{L}(\mathcal{H}^A)$). The joint probabilities of the preparation and measurement outcomes are given by $p(i,j|\{\rho^A_i\}_{i\in O}, \{E^A_j\}_{j\in Q}) ={\tr(\rho^A_i E^A_j)}$, $ \forall{i\in O}$, $\forall j\in Q  $. Assuming that the probability for the preparation event to belong to the subset $O'\subset O$ and the measurement event to belong to the subset $Q'\subset Q$ is nonzero, by locally discarding those cases in which the events do not belong to the respective subsets, we obtain two new operations connected to each other by the system $A$, whose joint probabilities are given by
\begin{gather}
p(i,j|\{\rho^A_i\}_{i\in O'}, \{E^A_j\}_{j\in Q'}) =\frac{\tr(\rho^A_i E^A_j)}{\sum_{l\in O', m\in Q'} \tr(\rho^A_l E^A_m)},  \label{newrule}
\end{gather}
for all $ {i\in O', j\in Q'}  $.

From Eq.~\eqref{newrule}, we see that two sets of preparation events described by operators $\{\rho^A_i\}_{i\in O'}$ and $\{\sigma^A_i\}_{i\in O'}$ yield the same probabilities if and only if their operators differ by an overall factor, $\sigma_i = \alpha\rho_i$, $\forall i \in O'$, $\alpha> 0$. The same holds for the sets of measurement events. We can therefore choose a normalization in order to dispose of the irrelevant degree of freedom. We define equivalent preparations and measurements to be described by sets of PS operators that satisfy the normalizations (we choose different normalizations for preparations and measurements to keep parallelism with the standard formalism, which can be seen as a special case of the new one):
\begin{gather}
\{\rho^A_i\}_{i\in O}, \hspace{0.2cm}\sum_{i\in O}\textrm{Tr}(\rho^A_i) = 1, \label{preparation}\\
\{E^A_j\}_{j\in Q}, \hspace{0.2cm}\sum_{j\in O}\textrm{Tr}(E^A_j) = d_A. \label{detection}
\end{gather}
Note that preparations are described just as before, but measurements are now more general as they do not have to satisfy $\sum_{j\in O}E^A_j=\id^A $.

Introducing the notation
\begin{gather}
\overline{\rho}^A \equiv \sum_{i\in O} \rho^A_i,
\end{gather}
\begin{gather}
 \overline{E}^A \equiv \sum_{j\in Q} E^A_j,
 \end{gather}
 we can write the main probability rule in the form
\begin{gather}
p(i,j|\{\rho^A_i\}_{i\in O}, \{E^A_j\}_{j\in Q}) =\frac{\tr(\rho^A_i E^A_j)}{\tr{(\overline{\rho}^A \overline{E}^A)}}, \hspace{0.2cm} \forall{i\in O, j\in Q}, \label{newrule2}
\end{gather}
for any preparation and measurement for which $\overline{E}^A\overline{\rho}^A\neq 0^A$. 

Unlike the standard approach to OPTs, here not all preparations and measurements defined over the same system are compatible---some of them are simply never found connected to each other. Equivalently, we can say that their connection results in the \textit{null event}. These are the preparations and measurements for which $\overline{E}^A\overline{\rho}^A=0^A$, where $0^A$ is the null operator on system $A$. The joint probabilities for the outcomes of such a pair of preparation and measurement can be defined to be all zero (i.e., no outcome occurs).

It is easily seen from Eq.~\eqref{newrule2} that the equivalence classes of preparation events, or \textit{states}, are now described by a pair of PS operators $(\rho^A;\overline{\rho}^A)$, where $\rho^A\leq \overline{\rho}^A$, $\tr(\overline{\rho}^A)=1$, while the equivalence classes of measurement events, or \textit{effects}, are described by a pair of PS operators $(E^A;\overline{E}^A)$, where $E^A\leq \overline{E}^A$, $\tr(\overline{E}^A)=d_A$. The joint probability rule for a pair of state and effect is given by Eq.~\eqref{state-effect}. 

Via Eq.~\eqref{state-effect}, states are real functions on effects and vice versa. However, the sets of states and effects are not closed under convex combinations (only some subsets of them are---those that correspond to the same $\overline{\rho}$ or $\overline{E}$). Even though we may conceive of the convex combinations of these functions, they generally do not correspond to events that can be obtained by local procedures in the preparation and measurement boxes. 

To see this, consider for example two deterministic preparations, each preparing one of two standard states with density operators $\overline{\rho}_1$ and  $\overline{\rho}_2$, $\overline{\rho}_1 \neq \overline{\rho}_2$, both of which can be assumed to have full support. In the above language of states described by two operators, these correspond to $(\overline{\rho}_1; \overline{\rho}_1)$ and $(\overline{\rho}_2; \overline{\rho}_2)$. Regarding them as functions on effects, imagine that we want to find a closed-box preparation procedure that yields a convex combination of these functions, e.g., $\frac{1}{3}(\overline{\rho}_1; \overline{\rho}_1) + \frac{2}{3}(\overline{\rho}_2; \overline{\rho}_2)$. Any preparation that we may perform inside a closed box (allowing both pre- and post-selection) is captured by standard preparations. The desired convex combination must therefore correspond to some state $(\rho;\overline{\rho})$. But it is easy to see that such a state does not exist. Indeed, the requirement that it yields the desired convex combination of probabilities with all effects of the form  $(E;\overline{E}=\id)$ implies that $\rho=\overline{\rho} = \frac{1}{3} \overline{\rho}_1 + \frac{2}{3} \overline{\rho}_2$. But then for effects $(E; \overline{E})$ with $\overline{E}\neq \id$, the probabilities would generally not respect the convex combination:
\begin{gather}
\frac{\tr\left( ( \frac{1}{3} \overline{\rho}_1 + \frac{2}{3} \overline{\rho}_2) E\right )}{ \tr\left((\frac{1}{3} \overline{\rho}_1 + \frac{2}{3} \overline{\rho}_2)\overline{E}\right)}\neq \frac{1}{3} \frac{\tr( \overline{\rho}_1 E )}{ \tr(\overline{\rho}_1\overline{E} )}+ \frac{2}{3} \frac{\tr( \overline{\rho}_2 E )}{ \tr(\overline{\rho}_2\overline{E} )}.
\end{gather}
Of course, we may simulate the desired convex combination by suitably post-selecting preparation and measurement events, but this requires joint post-selection, which is not achievable by separate closed-box procedures for the preparation and the measurement.

Note that a deterministic state, i.e., a state associated with the outcome of a single-outcome preparation, is described by a pair of identical density operators $(\overline{\rho}; \overline{\rho} )$. Because of this redundancy, we can parameterize the space of deterministic states by a single density operator, just like the space of deterministic states in the standard formulation of quantum theory. The probabilities for the outcomes of a measurement applied on a deterministic state $\overline{\rho}^A$ are given by
\begin{gather}
p(j|\{E^A_j\}_{j\in O}, \overline{\rho}^A) = \frac{\tr(\overline{\rho}^A E^A_j )}{\tr(\overline{\rho}^A\overline{E}^A)}, \hspace{0.2cm} \textrm{for} \hspace{0.1cm}  \tr (\overline{\rho}^A  \overline{E}^A)\neq 0,\notag\\
\hspace{1.1cm} =0,\hspace{0.2cm} \textrm{for} \hspace{0.1cm}  \tr (\overline{\rho}^A  \overline{E}^A)= 0,
\label{conditionalrule}
\end{gather}
which reduces to Born's rule in the special case $\overline{E}^A = \id^A$. In particular, the set of deterministic states can be regarded as real functions on effects. As we saw in the above example, this set is not closed under convex combinations, even though the operators by which we describe deterministic states form a convex set---the usual set of density operators. As functions of these operators, the probabilities for measurement outcomes are not linear, but we emphasize that this does not mean non-linearity in the {state} as defined in an operational sense.

The spaces of states and effects over a system $A$,  $\textrm{St}^A$ and $\textrm{Eff}^A$, can be equipped with a natural distance. Let $(\rho^A_1; \overline{\rho}^A_1)$ and $(\rho^A_2; \overline{\rho}^A_2)$ be two states in $\textrm{St}^A$. We can define the distance 
\begin{gather}
D_{\textrm{St}^A}\left((\rho^A_1; \overline{\rho}^A_1), (\rho^A_2; \overline{\rho}^A_2)\right)  = \notag\\
\textrm{Sup}_{(E^A;\overline{E}^A )\in \textrm{Eff}^A} \left|p \left(       (\rho^A_1; \overline{\rho}^A_1),           (E^A;\overline{E}^A )              \right) - p \left(             (\rho^A_2; \overline{\rho}^A_2)        , (E^A;\overline{E}^A )          \right) \right|\notag\\
\leq 1. \label{metric}
\end{gather}
The fact that $D_{\textrm{St}^A}$ is a distance function can be verified straightforwardly. (The distance on $\textrm{Eff}^A$ can be defined analogously.) However, note that $D_{\textrm{St}^A}$ is not a continuous function of $||\rho^A_1 - \rho^A_2||$ and $||\overline{\rho}^A_1 - \overline{\rho}^A_2||$, where $|| \cdot ||$ denotes the operator norm. For example, consider two states $(\rho^A_1; \overline{\rho}^A)$ and $(\rho^A_2; \overline{\rho}^A)$, $\overline{\rho}^A = \rho^A_1 +\rho^A_2$, associated with the two possible outcomes of a preparation. If $\rho^A_1$ and $\rho^A_2$ have different supports, the two states are maximally distant no matter how small $||\rho^A_1-\rho^A_2||$ may be, as long as it is nonzero. Indeed, in the case when one of $\rho^A_1$ or $\rho^A_2$ has support that is inside (but different from) the support of the other, say, $\textrm{Supp} (\rho^A_1) \subset \textrm{Supp} (\rho^A_2)$, we can achieve the maximum value of $1$ for the right-hand side of Eq.~\eqref{metric} by choosing an effect $(\overline{E}^A;\overline{E}^A)$ such that $\textrm{Supp}(\overline{E}^A)$ is orthogonal to $\textrm{Supp} (\rho^A_1)$ and has non-zero overlap with $\textrm{Supp} (\rho^A_2)$. In the case when none of $\rho^A_1$ or $\rho^A_2$ has support that is inside the support of the other, the same effect also yields the maximum value. The maximum distance between these states reflects the fact that there exists a measurement event that can occur together with one of the preparation events but not with the other.

As for preparations and measurements, one can define the equivalence classes of general operations and general events. Equivalent operations are described by a collection of CP maps with the normalization
\begin{gather}
\{\mathcal{M}^{A\rightarrow B}_j \}_{j\in O}, \hspace{0.2cm} \sum_{j\in O} \textrm{Tr}(\mathcal{M}^{A\rightarrow B}_j (\frac{\id^A}{d_A}) ) = 1, \label{genops}
\end{gather}
which reduces to the normalization of preparations and measurements in the respective limiting cases. Defining
\begin{gather}
\overline{\mathcal{M}}^{A\rightarrow B}= \sum_{i\in O}\mathcal{M}^{A\rightarrow B}_i,
\end{gather}
one sees that equivalent events, or \textit{transformations}, from $A$ to $B$ are described by pairs of CP maps
\begin{gather}
(\mathcal{M}^{A\rightarrow B}; \overline{\mathcal{M}}^{A\rightarrow B}) \label{transformations}
\end{gather}
with the properties
\begin{gather}
\mathcal{M}^{A\rightarrow B}(\rho^A) \leq \overline{\mathcal{M}}^{A\rightarrow B}(\rho^A), \forall \rho^A\geq 0, \\
\textrm{Tr}(\overline{\mathcal{M}}^{A\rightarrow B}(\frac{\id^A}{d_A}))=1.
\end{gather}

Although in OPTs it makes sense to think of an operation as a collection of transformations, here we choose to describe operations as collections of CP maps as above, which we find more natural in view of the intuition developed from the standard formulation.

Generalizing the case of preparations and measurements, two operations $\{\mathcal{N}^{B\rightarrow C}_j\}_{j\in Q}$ and $\{\mathcal{M}^{A\rightarrow B}_i\}_{i\in O}$ are not compatible, or their composition amounts to the null operation from $A$ to $C$, when $\overline{\mathcal{N}}^{B\rightarrow C} \circ \overline{\mathcal{M}}^{A\rightarrow B} = {0}^{A\rightarrow C}$, where ${0}^{A\rightarrow C}$ is the null CP map. The operation resulting from the sequential composition of two compatible operations, $\{\mathcal{L}^{A\rightarrow C}_{ij}\}_{i\in O,j\in Q} = \{\mathcal{N}^{B\rightarrow C}_j\}_{j\in Q}\circ \{\mathcal{M}^{A\rightarrow B}_i\}_{i\in O}$, has CP maps
 \begin{gather}
 \mathcal{L}^{A\rightarrow C}_{ji} =  \frac{ \mathcal{N}^{B\rightarrow C}_{j} \circ\mathcal{M}^{A\rightarrow B}_{i}} { \tr (\overline{\mathcal{N}}^{B\rightarrow C}\circ \overline{\mathcal{M}}^{A\rightarrow B}(\frac{\id^A}{d_A}))}, \hspace{0.2cm} i\in O, j \in Q.\label{seqcomp}
 \end{gather}

Upon learning or discarding of local information about the outcomes of an operation, its description gets updated. To derive the most general update rule, it is convenient to model the classical variable describing the outcome of an operation by a set of (standard) orthonormal pure pointer `states' $\{|i\rangle^C\}_{i\in O}$ on a pointer system $C$. An operation $\{\mathcal{M}_i^{A\rightarrow B}\}_{i\in O}$ can then be thought of as a two-step process, the first step being the single-outcome operation
\begin{gather}
\mathcal{M}^{A\rightarrow BC}=\sum_{i\in O} \mathcal{M}_i^{A\rightarrow B}\otimes |i\rangle\langle i|^C, \label{instro}
\end{gather}
and the second one being a standard von Neumann measurement of the system $C$ in the pointer basis. Without loss of generality, we can imagine that the outcome of the measurement is stored in another pointer system, so for an experimenter who has not looked at the information about the outcome of the operation, the experiment can be described by the first stage only (this is nothing but the model of a standard quantum instrument \cite{instrument}, trivially extended to the more general type of operations we consider). Any process of learning or discarding of information about the outcome of the operation can be described by a classical operation on the pointer system. This most generally corresponds to a diagnonal CP map, followed by a renormalization of the overall operation, which results in an operation of a similar form to \eqref{instro}, but with the pointer states possibly running over a different set. Let $C'$ describe the (possibly different) pointer system after this operation, with pointer basis $\{|j\rangle^{C'} \}_{j\in Q}$. The diagonal CP map describing the transformation of the pointer has the form
\begin{gather}
\mathcal{M}^{C\rightarrow C'} (\cdot) = \sum_{i\in O, j\in Q} T(j,i) |j \rangle^{C'}\langle i|^C (\cdot) |i \rangle^{C}\langle j|^{C'} ,
\end{gather}
 where
 \begin{gather}
 T(j,i)\geq 0, \hspace{0.2cm}\forall i\in O, \forall j \in Q, \hspace{0.2cm}\sum_{j\in Q} T(j,i)\leq 1, \forall i\in O.
 \end{gather}
 After renormalization, this gives rise to the updated overall operation
\begin{gather}
\mathcal{M}^{A\rightarrow BC'}=\frac{\sum_{j\in Q} \sum_{i\in O} T(j,i)\mathcal{M}_i^{A\rightarrow B}\otimes |j\rangle\langle j|^{C'}}{\sum_{j\in Q} \sum_{i\in O} T(j,i)\tr(\mathcal{M}_i^{A\rightarrow B}(\frac{\id^A}{d_A}))}.\label{instro2}
\end{gather}
From this, we infer the general update rule of the operation on the original systems:
\begin{gather}
\{ \mathcal{M}^{A\rightarrow B}_i\}_{i\in O} \rightarrow \{ \mathcal{M}'^{A\rightarrow B}_j\}_{j\in Q},
\end{gather}
where
\begin{gather}\label{operationupdate}
\mathcal{M}'^{A\rightarrow B}_j=  \frac{\sum_{i\in O} T(j,i)\mathcal{M}_i^{A\rightarrow B}} {\sum_{j\in Q} \sum_{i\in O} T(j,i)\tr(\mathcal{M}_i^{A\rightarrow B}(\frac{\id^A}{d_A}))}\hspace{0.2cm}.
\end{gather}

It is worth noting a couple of special cases that will be used later. The case of completely discarding the information about the outcome of the operation corresponds to $Q=\{e\}$ being a singleton set, and $T(e,i)=1$, $\forall i\in O$. This leads to the fully coarse-grained deterministic operation $\{\overline{\mathcal{M}}^{A\rightarrow B}\}$. On the other hand, the case in which the outcome of the operation is learned to belong to a specific subset, $O'\subset O$, corresponds to $Q=O'$ and $T(i,i) = p>0$ for $i\in O'$, and $T(j,i) = 0$ in all other cases. The latter gives us a prescription of how to obtain any operation in the new formalism starting from a standard operation (one whose CP maps sum up to a CPTP map) and using post-selection. It is important to emphasize, however, that the theory does not attribute any special status to those operations that satisfy the standard trace-preserving condition.

To summarize, the generalized formulation is defined by the following rules:\\

(i) Systems are associated with Hilbert spaces, and an operation from a given input to a given output system is described by a collection of CP maps from the space of operators over the input Hilbert space to the space of operators over the output Hilbert space, with the normalization \eqref{genops}.\\

(ii) When a preparation is connected to a measurement, the joint probabilities for their outcomes are given by Eq.~\eqref{newrule2}. Equivalently, two operations connected sequentially yield a new operation according to \eqref{seqcomp}, and CP maps from the trivial system to itself are interpreted as probabilities.\\

(iii) Upon learning or discarding of information, the description of an operation is updated according to Eq.~\eqref{operationupdate}.\\

Even though we have formulated the theory for finite dimensions, we expect that it can be extended to infinite dimensions with suitable modifications of the representation conventions. 

We have described all possible circuits and the probabilities for their outcomes in the generalized formulation of quantum theory using the mathematical language of Hilbert spaces. An interesting question is to find a set of operational principles from which this formulation can be derived, similarly to the way this has been done for the standard formulation \cite{CDP,Hardy2,infounit}.

\subsection{Time reversal and general symmetries: proof of the main theorem}

Since we have allowed operations to be defined by both pre- and post-selection, one can expect that the theory should be symmetric under time reversal in some sense. This is because the events that constitute a valid operation in one direction of time constitute a valid operation in the other, while the probabilities of events conditional on specific information are independent of the direction of time. Here, we discuss in detail the question of time reversal along with general symmetry transformations.

Under time reversal $\mathcal{T}$, every operation from $A$ to $B$ is expected to be seen as a valid operation from $B$ to $A$, such that the probabilities of any circuit when calculated in the opposite direction under this transformation remain the same. This by itself, however, does not define time reversal. Indeed, we will see that the above theory permits infinitely many transformations with this property. Time reversal is a specific, physically motivated transformation, which is not implicit in the formalism. The simplest example of a transformation that satisfies the general requirement above is the following. For every CP map $\mathcal{M}^{A\rightarrow B}$, which can be written in the Kraus form $\mathcal{M}^{A\rightarrow B} (\cdot) = \sum_{\alpha=1}^{d_A d_B }  K_{\alpha} (\cdot) K_{\alpha}^{\dagger}$, where $K_{\alpha}: \mathcal{H}^A\rightarrow \mathcal{H}^B$ are linear maps \cite{Kraus}, we could define the `time-reversed' image as the CP map $\mathcal{M}^{\dagger B\rightarrow A} (\cdot) = \frac{d_B}{d_A}\sum_{\alpha=1}^{d_A d_B} K^{\dagger}_{\alpha} (\cdot) K_{\alpha}$, where $K^{\dagger}_{\alpha}:\mathcal{H}^B\rightarrow \mathcal{H}^A$. This definition is basis-independent, and it simply amounts to reading a circuit in the opposite direction by regarding the operators of preparations as operators of measurements up to a (dimension-dependent) constant factor, and vice versa. More precisely, a preparation $\{\rho^A_i\}_{i\in O}$ is seen as a measurement $\{d_A\rho^A_i\}_{i\in O}$, and a measurement $\{E^A_j\}_{j\in Q}$ as a preparation $\{\frac{1}{d_A}E^A_j\}_{j\in Q}$. (At the level of the underlying Hilbert space, this is equivalent to interchanging vectors $|\psi\rangle\in \mathcal{H}$ with their canonical duals  $\langle \psi| \in \mathcal{H}^*$ \cite{AV}, up to a factor.) Using the cyclic invariance of the trace, one can easily see that the probabilities of a circuit remain invariant under this transformation. The new states and measurements correspond to the so called \textit{retrodictive} states and measurements \cite{Watanabe, Barnett, LeiferSpekkens}.

The problem with this definition arises when one goes beyond the mere OPT and makes a connection to concepts such as energy, momentum, or spin. The latter are not part of the OPT \textit{per se}, but are the subject of the theory that describes the dynamics of physical systems, which we will refer to as the \textit{mechanics} (an OPT says what the possible operations are, but not what operations will arise in specific circumstances). According to our present understanding of the laws of mechanics, an isolated quantum system undergoes unitary evolution in time driven by a Hamiltonian generator, which is the operator of energy. Quantum states are described in terms of physical variables such as momentum, or spin. Under time reversal, these variables transform in a specific way (e.g., energy remains invariant, while momentum and spin change sign) and this determines the notion of time reversal in a physical sense. As shown by Wigner \cite{Wigner} (see also Schwinger \cite{Schwinger}), these considerations imply that time reversal must be described by an anti-unitary transformation at the Hilbert-space level (see below). The above transformation, however, does not correspond to an anti-unitary transformation on the Hilbert space. If we assume, as is the current understanding of quantum mechanics, that the states of an isolated system evolve forward in time according to the Schr\"{o}dinger equation driven by a given Hamiltonian with a positive energy spectrum, then the retrodictive states would evolve backward in time driven by the negative of the original Hamiltonian, which would have a negative energy spectrum.

To understand the issue of time reversal, let us have a closer look at the relation between preparation and measurement events, and their representations. The operators $\{\rho^A_i\}_{i\in O}$ by which we describe a preparation can be thought of as elements of the real vector space $\mathcal{V}^{A}$ of Hermitian operators over $\mathcal{H}^A$ (strictly speaking, a preparation is a collection of CP maps, $\{\rho^{I\rightarrow A}_i\}_{i\in O}$, which are elements of the real vector space of linear maps from $\mathbb{R}$ to $\mathcal{V}^{A}$, which is naturally isomorphic to $\mathcal{V}^{A}$). Measurements can be similarly thought of as described by collections of vectors, but in the dual vector space, $\mathcal{V}^{A^*}$. This dual vector space is isomorphic to $\mathcal{V}^{A}$ if $\mathcal{V}^{A}$ is finite-dimensional. Let us denote the vectors in the dual space by $E^{A^*}$. The pairing between elements of the two vectors spaces yields a real number: $(E^{A^*},\rho^A) \in \mathbb{R}$, $\forall E^{A^*}\in \mathcal{V}^{A^*}, \forall \rho^A \in \mathcal{V}^{A}$, which we write as $(E^{A^*},\rho^A)= \tr[\rho^A E^A ]$, where $E^A\in \mathcal{V}^{A}$ corresponds (via an isomorphism) to $E^{A^*}$. Note, however, that prior to choosing this representation, there is no natural isomorphism between the vector space $\mathcal{V}^{A}$ and its dual $\mathcal{V}^{A^*}$. Every non-degenerate bilinear form $\langle \cdot, \cdot \rangle : \mathcal{V}^{A} \times \mathcal{V}^{A} \rightarrow \mathbb{R}$ gives rise to an isomorphism. Our representation is based on the particular choice of bilinear form  $\langle\rho^A, \sigma^A\rangle \equiv \tr[\rho^A\sigma^A]\in \mathbb{R}$. 
This bilinear form is an inner product (the Hilbert-Schmidt inner product restricted to the subspace of Hermitian operators). It induces a `canonical' isomorphism between the two vector spaces, $E^{A^*}\leftrightarrow E$, $E^{A^*}\in \mathcal{V}^{{A^*}}$, $E\in \mathcal{V}^{{A}}$, given via $(E^{A^*}, \rho^A) = \langle\rho^A, E^A\rangle = \tr[\rho^A E^A ]$. This canonical isomorphism merely corresponds to a choice of representation for the pairing between dual vectors, and need not have any physical meaning. (The physically nontrivial aspect of this picture is that the vector spaces can be realized as the spaces of Hermitian operators over a complex Hilbert space of a given dimension.) As we will show below, time reversal may also define an isomorphism between the two dual vector spaces (though not necessarily, since the general correspondence is between states and effects, and these are described by pairs of vectors) because under time reversal measurements are mapped onto preparations, and vice versa. However, this cannot be the canonical isomorphism arising from the standard choice of bilinear form because of the physical considerations noted earlier. The retrodictive states and measurements arise exactly from the canonical isomorphism, up to a constant factor. 

Under time reversal $\mathcal{T}$, every set of vectors $\{\rho^A_i\}_{i\in O}$ in $ \mathcal{V}^{{A}}$ corresponding to a valid preparation would become a specific set of vectors $\{ F^{A^*}_i\}_{i\in O}$ in $ \mathcal{V}^{{A^*}}$ corresponding to a valid measurement, which is the \textit{time-reversed image} of $\{\rho^A_i\}_{i\in O}$. 
Using the canonical isomorphism, the time-reversed image of the preparation will be described by a set of measurement operators $\{ F^{A}_i\}_{i\in O}$ in $\mathcal{V}^{{A}}$ (Fig. 3). In a similar way, the operators $\{E^A_j\}_{j\in Q}$ describing a measurement get mapped onto a set of operators $\{\sigma^A_j\}_{j\in Q}$ describing a preparation, which is the time-reversed image of the measurement. Under this interchange, which must be invertible, the probabilities \eqref{newrule2} (or equivalently, \eqref{state-effect}) must remain the same. The latter requirement means that $\mathcal{T}$ is described by a bijection between states $(\rho^A;\overline{\rho}^A)$, $\rho^A, \overline{\rho}^A\in \mathcal{V}^A$ and effects $(E^{A^*};\overline{E}^{A^*})$, $E^{A^*}, \overline{E}^{A^*} \in \mathcal{V}^{A^*}$ (where we represent the effects via the canonical isomorphism as $(E^{A};\overline{E}^{A})$, $E^{A}, \overline{E}^{A} \in \mathcal{V}^{A}$). This does not, however, imply that time reversal should be realized by a bijection between $\mathcal{V}^A$ and $\mathcal{V}^{A^*}$ applied independently on each vector in the pair $(\rho^A;\overline{\rho}^A)$ (or $(E^{A^*};\overline{E}^{A^*})$). Indeed, below we completely characterize the transformations that preserve the probabilities and show that more general transformations are possible. 

As pointed out in the main text, since states and effects are different objects, a symmetry transformation has to be defined by its action on both the space of states and the space of effects, $\mathcal{S}^A:\textrm{St}^A\times \textrm{Eff}^A\rightarrow \textrm{St}^A\times \textrm{Eff}^A$. Symmetry transformations of \textit{type} $I$, such as spatial rotation, transform states into states and effects into effects. Symmetry transformations of \textit{type} $II$, such as time reversal or any combination of time reversal with a symmetry transformation of \textit{type} $I$, transform states into effects and effects into states. In the case of transformations of \textit{type} $I$, the transformation $\mathcal{S}^A$ can be thought of as consisting of two transformations, $(\mathcal{S}^A_{s\rightarrow s}, \mathcal{S}^A_{e\rightarrow e})$, where $\mathcal{S}^A_{s\rightarrow s}: \textrm{St}^A\rightarrow \textrm{St}^A$, $\mathcal{S}^A_{e\rightarrow e}: \textrm{Eff}^A\rightarrow \textrm{Eff}^A$, and in the case of \textit{type} $II$, it can be thought of as consisting of two transformations, $(\mathcal{S}^A_{s\rightarrow e}, \mathcal{S}^A_{e\rightarrow s})$, where $\mathcal{S}^A_{s\rightarrow e}: \textrm{St}^A\rightarrow \textrm{Eff}^A$, $\mathcal{S}^A_{e\rightarrow s}: \textrm{Eff}^A\rightarrow \textrm{St}^A$. By representing effects in terms of pairs of vectors in $\mathcal{V}^A$ via the canonical isomorphism, each of these transformations can be represented as a transformation on the space of pairs of PS operators on $\mathcal{H}^A$. We denote these representations by  $\hat{S}^A_{s\rightarrow s}$, $\hat{S}^A_{e\rightarrow e}$, $\hat{S}^A_{s\rightarrow e}$, $\hat{S}^A_{e\rightarrow s}$, respectively. The possible form of these symmetry transformations is given by our main theorem, which we prove next. (We drop the superscript $A$ for the proof of the theorem.)

\textit{Proof of Theorem}. Since we are interested primarily in time reversal, we will exhibit the proof for transformations of \textit{type} $II$. The case of \textit{type} $I$ is analogous. We will make use of the way operations get updated upon learning or discarding of information (Eq.~\eqref{operationupdate}), which must be independent of the symmetry transformation. First, observe that the case of complete coarse graining implies that two states have the same $\overline{\rho}$ if and only if their images under the symmetry transformation have the same $\overline{F}$. (The same holds for measurements and their images.) 
Consider then two states $(\rho_1; \overline{\rho})$ and $(\rho_2; \overline{\rho})$ whose images are $(F_1; \overline{F})$ and $(F_2; \overline{F})$, respectively. Let us take any state $(q\rho_1+(1-q)\rho_2; \overline{\rho})$, $0\leq q \leq 1$. From formula \eqref{state-effect}, we see that the joint probability of this state with any effect $(E; \overline{E})$ is $p\left((q\rho_1+(1-q)\rho_2; \overline{\rho}), (E; \overline{E}) \right)  = q p \left((\rho_1; \overline{\rho}), (E; \overline{E}) \right)  + (1-q) p\left((\rho_2; \overline{\rho}), (E; \overline{E}) \right) $. Similarly, consider the effect $(qF_1+(1-q)F_2; \overline{F})$, with the same $q$. It must yield the probabilities  $p\left((\sigma; \overline{\sigma}), (qF_1+(1-q)F_2; \overline{F}) \right)  = q p \left((\sigma; \overline{\sigma}), (F_1; \overline{F}) \right)  + (1-q) p \left((\sigma; \overline{\sigma}), (F_2; \overline{F}) \right)$ when paired with a state $(\sigma; \overline{\sigma})$. But when $(\sigma; \overline{\sigma})$ is the image of $(E;\overline{E})$, the probabilities in the first case must be equal to the corresponding probabilities in the second case. Since a state (effect) is completely characterized by its joint probabilities with all possible effects (states), we conclude that $(qF_1+(1-q)F_2; \overline{F})$ must be the image of $(q\rho_1+(1-q)\rho_2; \overline{\rho})$. In other words, for every fixed $\overline{\rho}$, $\hat{S}_{s\rightarrow e}$ transforms the first operator in $( \rho; \overline{\rho})$ by a (positive) linear map, possibly dependent on $\overline{\rho}$, which we will denote by $\hat{\tau}_{\overline{\rho}}$. This linear map can be assumed defined on the subspace of Hermitian operators with support in the support of $\overline{\rho}$. Consider now the update rule \eqref{operationupdate} in the case of learning the outcome of an operation (a special case of learning that the outcome belongs to a subset, which was discussed earlier). It implies that two states have proportional $\rho$ (differing by an overall factor) if and only if their images have proportional $F$. This means that $\hat{\tau}_{\overline{\rho}} (\rho) = f(\overline{\rho}) \hat{\tau}_{\frac{\id}{d}} (\rho) \equiv f(\overline{\rho}) \hat{\tau}(\rho)$, for all $\rho$ in the domain of $\hat{\tau}_{\overline{\rho}}$, where $\hat{\tau}\equiv \hat{\tau}_{\frac{\id}{d}}$ is a positive linear map defined on the whole space of Hermitian operators over $\mathcal{H}$. But since every deterministic state must be mapped onto a deterministic effect, we have $\hat{\tau}_{\overline{\rho}} (\overline{\rho})=\overline{F}$, and hence $  f(\overline{\rho})\tr[ \hat{\tau}(\overline{\rho})]  =   \tr[\hat{\tau}_{\overline{\rho}} (\overline{\rho})]= d$, which implies $f(\overline{\rho}) = {d}/{\tr[\hat{\tau}( \overline{\rho}) ]} $. We thus obtain
\begin{gather}
\hat{S}_{s\rightarrow e} (\rho;\overline{\rho}) = (d \frac{\hat{\tau} ( \rho) }{\tr [\hat{\tau}(\overline{\rho}   ) ] };  d \frac{\hat{\tau} (\overline{\rho} ) }{\tr [\hat{\tau}(\overline{\rho}   ) ] }   ).
\end{gather}
Since $\hat{S}_{s\rightarrow e} $ is a bijection, $\hat{\tau}$ must map the cone of PS operators over $\mathcal{H}$ onto itself. This means \cite{Wolf} that $\hat{\tau}$ is either of the form $\hat{\tau} (\rho) = S\rho S^{\dagger}$ or of the form $\hat{\tau} (\rho) = S\rho^T S^{\dagger}$, where $S$ is invertible, which corresponds to Eqs.~\eqref{Th1} or \eqref{Th2}. An analogous argument applied to the transformation of effects yields Eqs.~\eqref{Th3} and \eqref{Th4}. \qed

\textit{Note.} The operator $S$ depends on the transposition basis. The basis can be chosen arbitrarily by redefining $S$. For involutions, in the case of Eqs.~\eqref{th1}, \eqref{th2}, $S$ satisfies $S=\propto S^{-1}$, in the case of Eqs.~\eqref{th3}, \eqref{th4}, $S$ satisfies $S=\propto {{S^*}^{-1}}$, where $^*$ denotes complex conjugation in the basis of the transposition, in the case of Eqs.~\eqref{Th1}, \eqref{Th2}, $S$ satisfies $S\propto S^{\dagger}$, and in the case of Eqs.~\eqref{Th3}, \eqref{Th4}, $S$ satisfies $S=\pm S^{T}$. This follows straightforwardly from the requirement that applying the transformation twice maps every state and effect onto itself.

Let us assume, as in standard quantum mechanics, that isolated systems evolve in time unitarily according to the Schr\"{o}dinger equation driven by some Hamiltonian, and let us assume, following Wigner \cite{Wigner}, that the same kind of evolution should take place under time reversal, driven by a Hamiltonian with the same spectrum (since energy should not change under time reversal). Let a general transformation $(\mathcal{M}^{A\rightarrow B}; \overline{\mathcal{M}}^{A\rightarrow B} )$, where $\mathcal{M}^{A\rightarrow B} (\cdot)= \sum_{\alpha=1}^{d^Ad^B} K_{\alpha} (\cdot) K^{\dagger}_{\alpha}$, $\overline{\mathcal{M}}^{A\rightarrow B} (\cdot)= \sum_{\alpha=1}^{d^Ad^B} \overline{K}_{\alpha} (\cdot) \overline{K}^{\dagger}_{\alpha}$, be transformed under time reversal $\mathcal{T}$ as
\begin{gather}
(\mathcal{M}^{A\rightarrow B}; \overline{\mathcal{M}}^{A\rightarrow B} ) \overset{\mathcal{T}}\rightarrow (\tilde{\mathcal{M}}^{B\rightarrow A}; \overline{\tilde{\mathcal{M}}}^{B\rightarrow A} ),
\end{gather}
\begin{gather}
\tilde{\mathcal{M}}^{B\rightarrow A}  (\cdot)= \sum_{\alpha=1}^{d^Ad^B} \tilde{K}_{\alpha}(\cdot) \tilde{K}^{\dagger}_{\alpha},\\
\overline{\tilde{\mathcal{M}}}^{B\rightarrow A}  (\cdot)= \sum_{\alpha=1}^{d^Ad^B} \overline{\tilde{K}}_{\alpha}(\cdot) \overline{\tilde{K}}^{\dagger}_{\alpha}.
\end{gather}
If $\mathcal{T}$ transforms states and effects as in Eqs.~\eqref{Th1} and \eqref{Th2} (with specific transposition bases and specific operators $S^A$ and $S^B$ for the respective systems), from the requirement that the probability for a sequence of a state, a transformation, and an effect, remains invariant, we find
\begin{gather}
\tilde{K}_{\alpha} =   (S^B K_{\alpha} {S^A}^{-1})^{\dagger} /\lambda,
\end{gather}
\begin{gather}
\overline{\tilde{K}}_{\alpha} =   (S^B \overline{K}_{\alpha} {S^A}^{-1})^{\dagger} /\lambda,
\end{gather}
for all $\alpha = 1 \dots d^Ad^B$, with $\lambda$ such that
\begin{gather}
\tr(\sum_{\alpha}^{d^Ad^B} \overline{\tilde{K}}_{\alpha}^{\dagger} \overline{\tilde{K}}_{\alpha}) = d^B. \label{lambda}
\end{gather}

However, if in the case when $A$ is of the same kind as $B$ a unitary transformation ($K_{\alpha} = \overline{K_{\alpha}} =\delta_{\alpha,1}U$, $U^{\dagger}U = \id$) could be mapped onto a unitary ($\tilde{K}_{\alpha} = \overline{\tilde{K}_{\alpha}} =\delta_{\alpha,1}\tilde{U}$, $\tilde{U}^{\dagger}\tilde{U} = \id$), where $\tilde{U}=(S U S^{-1})^{\dagger}$ has the same spectrum as $U$, there would have to exist a unitary $W$ (from the input to the output system of $U$) such that $W {S^{-1}}^{\dagger} U^{\dagger} S^{\dagger}W^{\dagger}=U$. But the left-hand side of the last expression is a similarity transformation of $U^{\dagger}$ which preserves the spectrum, so this is only possible if $U$ has a real spectrum (consisting of $+1$'s and $-1$'s), which does not permit nontrivial continuous unitary evolution in time.

Thus, the only possibility compatible with the known quantum mechanics is that time reversal is described by a transformation of the form \eqref{Th3} and \eqref{Th4}. In such a case, we find
 \begin{gather}
 \tilde{K}_{\alpha} =   ({S}^B K^*_{\alpha} {S^A}^{-1})^{\dagger} /\lambda, \label{Th5}
 \end{gather}
 \begin{gather}
 \overline{\tilde{K}}_{\alpha} =   ({S}^B \overline{K}^*_{\alpha}  {S^A}^{-1})^{\dagger} /\lambda, \label{Th6}
 \end{gather}
 for all $\alpha = 1 \dots d^Ad^B$, where $^*$ denotes complex conjugation in the joint basis in which the transpositions for $A$ and $B$ are defined in Eqs.~\eqref{Th3} and \eqref{Th4}, and $\lambda$ ensures the normalization \eqref{lambda}. In this case, for the image of a unitary operation we obtain $\tilde{U}=(S U^* {S}^{-1})^{\dagger}$. If $S=V$ is unitary, $\tilde{U}=VU^TV^{\dagger}$ would be unitary and it would have the same spectrum as $U$, since $U^T $ has the same spectrum as $U$. Note that it is not necessary that $S$ be unitary in order for $\tilde{U}$ to satisfy this property. If $S$ has a polar decomposition $S=VM$, $M\geq 0$, where $V$ is unitary and $M$ commutes with $U^T$ (or, equivalently, with the transpose of the Hamiltonian generator of $U$), then the requirement is still satisfied. However, if we further demand that time reversal satisfies the above requirement for any Hamiltonian generator, then $S$ must be unitary. The standard notion of time reversal, as understood at present, corresponds to this case, although it is formulated as a map from the state space to itself. Since we are generalizing the standard formulation of quantum theory, it is in principle conceivable that in some regimes the laws of mechanics may not obey Schr\"{o}dinger's equation, which was used in the above argument. It is reasonable to assume, however, that any generalized notion of time reversal would be of the kind \eqref{Th3}, \eqref{Th4}, (equivalently, \eqref{Th5}, \eqref{Th6}) so that it would reduce continuously to the standard one in the regimes of standard quantum mechanics. 
 
Since time reversal is a reflection, its transformation of states and effects is expected to be an involution. This means that $S=S^T$ or $S=-S^T$. When $S$ is unitary, these two cases correspond to the form of time reversal for bosons and fermions, respectively \cite{Wick}. Note that the bosonic time reversal is an involution also at the level of the Hilbert space $\mathcal{H}$, but the fermionic one is not since applying it twice yields an overall minus sign. The minus sign disappears at the level of the operators by means of which we describe states and effects, but its existence at the Hilbert-space level is one way of arguing that there has to be a spin superselection rule \cite{Wick}.

It is also interesting to note that when $S$ is unitary, time reversal corresponds to an isomorphism between the underlying Hilbert space $\mathcal{H}$ and its dual $\mathcal{H^*}$, which is \textit{linear}. The standardly claimed anti-linearity of time reversal arises from the representation of the vectors in $\mathcal{H^*}$ by vectors in $\mathcal{H}$ via the canonical anti-linear isomorphism between the two spaces. 


The most general possible form of time reversal \eqref{Th5}, \eqref{Th6} on an arbitrary transformation was obtained from the requirement that the probabilities for a preparation, followed by a general operation, followed by a measurement, remain the same under time reversal. One can easily see that this guarantees that the probabilities remain invariant for general circuits, since any circuit can be `foliated' into global time steps where at each step a single operation is applied from a given composite input system to a given composite output system (this can be achieved by padding operations with additional sequences of identity operations where necessary).  The joint probabilities of a circuit consisting of a preparation $\{\rho^{A_0}_{i_0}\}_{i_0\in Q_0}$, followed by a sequence of operations $\{ \mathcal{M}^{A_{n-1}\rightarrow A_{n}}_j\}_{i_n\in Q_n}$, $n=1,...,N-1$, and then by a measurement $\{E^{A_N}_{i_N}\}_{i_N\in Q_N}$, are given by
\begin{gather}
p(i_0,i_1,\cdots, i_N| \{\rho^{A_0}_{i_0}\}_{i_0\in Q_0}, \{\mathcal{M}^{A_0\rightarrow A_1}_{i_1}\}_{i_1\in Q_1}  ,\cdots, \{E^{A_N}_{i_N}\}_{i_N\in Q_N}) \notag\\= \frac{\tr(E^{A_N}_{i_N} \mathcal{M}^{A_{N-1}\rightarrow A_{N}}_{i_{N-1}}(\cdots \mathcal{M}^{A_0\rightarrow A_1}_{i_1}(\rho^{A_0}_{i_0})) )}{\tr(\overline{E}^{A_N} \overline{\mathcal{M}}^{A_{N-1}\rightarrow A_{N}}(\cdots \overline{\mathcal{M}}^{A_0\rightarrow A_1}(\overline{\rho}^{A_0})) )}, \label{straight}
\end{gather}
and the fact that the probabilities remain invariant under the transformation given by Eqs.~\eqref{Th5} and \eqref{Th6} can be verified by expanding each CP map in its Kraus form and using the invariance of the trace under cyclic permutations and transposition.

\section{Supplementary Methods}

\subsection{Sequential and parallel composition of operations}

Operations can be composed in sequence and in parallel to form new operations \cite{Coecke, CDP2}. 

Two operations can be composed in sequence if the output system of the first one is of the same type as the input system of the second one: $\{\mathcal{N}^{B\rightarrow C}_j\}_{j\in Q}\circ \{\mathcal{M}^{A\rightarrow B}_i\}_{i\in O}=  \{\mathcal{N}^{B\rightarrow C}_j\circ \mathcal{M}^{A\rightarrow B}_i\}_{i\in O,j\in Q} \equiv\{\mathcal{L}^{A\rightarrow C}_k\}_{k\in O\times Q}$. For every system $A$, one defines the identity operation $\mathcal{I}^{A\rightarrow A}$, which has a single outcome and has the property $\{\mathcal{M}^{A\rightarrow B}_i\}_{i\in O}\circ \mathcal{I}^{A\rightarrow A} = \{\mathcal{M}^{A\rightarrow B}_i\}_{i\in O}$ and $\mathcal{I}^{A\rightarrow A}\circ \{\mathcal{N}^{B\rightarrow A}_j\}_{j\in Q}= \{\mathcal{N}^{B\rightarrow A}_j\}_{j\in Q}$, for all $\{\mathcal{M}^{A\rightarrow B}_i\}_{i\in O}$ and $\{\mathcal{N}^{B\rightarrow A}_j\}_{j\in Q}$.

Given a pair of systems $A$ and $B$, one can define the composite system $AB$. The parallel composition of two operations $\{\mathcal{M}^{A\rightarrow B}_i\}_{i\in O}$ and $\{\mathcal{N}^{C\rightarrow D}_j\}_{j\in Q}$ is a new operation $\{\mathcal{M}^{A\rightarrow B}_i\}_{i\in O}\otimes \{\mathcal{N}^{C\rightarrow D}_j\}_{j\in Q}=\{\mathcal{M}^{A\rightarrow B}_i\otimes \mathcal{N}^{C\rightarrow D}_j\}_{i\in O,j\in Q}\equiv \{\mathcal{L}^{AC\rightarrow BD}_k\}_{k\in O\times Q}$ from the composite system $AC$ to the composite system $BD$. 

As circuits have no open wires, the composition of operations in a circuit amounts to an operation from the trivial system to itself.

\subsection{Canonical representation of states and effects in the standard formulation of quantum theory}

Every CP map from system $A$ to system $B$ can be written in the (generally non-unique) Kraus form \cite{Kraus}
$\mathcal{M}^{A\rightarrow B} (\cdot) = \sum_{\alpha=1}^{d_Ad_B} K_{\alpha} (\cdot) K_{\alpha}^{\dagger}$,  where $\{K_{\alpha}\}_{\alpha=1}^{d_Ad_B}$, $K_{\alpha}: \mathcal{H}^A\rightarrow \mathcal{H}^B$, are linear maps, known as Kraus operators. The condition that the CP maps $\{\mathcal{M}^{A\rightarrow B}_i\}_{i\in O}$ associated with all outcomes of a standard quantum operation have a sum $\sum_{i\in O} \mathcal{M}^{A\rightarrow B}_i = \overline{\mathcal{M}}^{A\rightarrow B}$ that is a CP and trace-preserving (CPTP) map is equivalent to the constraint $\sum_{i\in O}\sum_{\alpha_i=1}^{d_Ad_B}  K_{\alpha_i}^{\dagger}K_{\alpha_i}= \id^A$, where $\{K_{\alpha_i}\}_{\alpha_i=1}^{d_Ad_B}$ are Kraus operators for $\mathcal{M}^{A\rightarrow B}_i$. The trivial system $I$ corresponds to the 1-dimensional Hilbert space $\mathbb{C}^1$. States are thus CP maps of the form $\rho^{I\rightarrow A} (\cdot)= \sum_{\alpha=1}^{d_A} |\psi_{\alpha}\rangle (\cdot) \langle \psi_{\alpha}|$, $|\psi_{\alpha}\rangle \in \mathcal{H}^A$. They are isomorphic to positive semidefinite (PS) operators, $\rho^{I\rightarrow A}\leftrightarrow \rho^A= \sum_{\alpha=1}^{d_A} |\psi_{\alpha}\rangle \langle \psi_{\alpha}|^A \in {\mathcal{L}(\mathcal{H}^A)}$, and this is how they are represented. In particular, a preparation is described by a set of PS operators $\{\rho^A_i\}_{i\in O}$, such that $\sum_{i\in O}\tr(\rho^A_i) = 1$. Effects are CP maps of the form $E^{A\rightarrow I}(\cdot)= \sum_{\alpha=1}^{d_A} \langle\phi_{\alpha}| (\cdot)|\phi_{\alpha} \rangle$, $|\phi_{\alpha} \rangle \in \mathcal{H}^A$. They are also isomorphic to PS operators, $E^{I\rightarrow A}\leftrightarrow E^A= \sum_{\alpha=1}^{d_A} |\phi_{\alpha}\rangle \langle \phi_{\alpha}|^A \in\mathcal{L}(\mathcal{H}^A)$, and this is how they are represented. Here the trace-preserving condition means that a measurement is described by a set of PS operators $\{E^A_j\}_{j\in Q}$ that form a positive operator-valued measure (POVM), i.e., $\sum_{j\in Q} E^A_j=\id^A$. In this representation, the joint probabilities of states and effects are given by $p(\rho^{I\rightarrow A}_i, E^{A\rightarrow I}_j) = E_j^{A\rightarrow I} \circ \rho_i^{I\rightarrow A} = \tr(\rho^A_i E^A_j)$.

\subsection{The concept of operation}

\subsubsection{The {closed-box} assumption}\label{closedboxassumption}

The concept of circuit formalizes the intuitive notion of information exchange. Notice that a circuit does not represent merely a sequence of applications of physical devices as one may commonly understand this. Indeed, we can construct a sequence in which the device applied at a given step is chosen based on information about the outcomes of previously applied devices according to some protocol. In such a case, the joint probabilities for the outcomes of the sequence would generally depend on the protocol, whereas a circuit is defined to have unique probabilities. The idea of a circuit is that it provides a complete picture of the information exchange responsible for the correlations between the set of events it describes---the wires in a circuit are assumed to represent all systems through which the correlations between the possible events in the boxes arise. The above example would not correspond to a valid circuit because it does not take into account all existing means of information exchange between the events. If we have a scenario in which the device at a given step is selected using information about past outcomes, in the language of circuits this would have to be described by an operation acting on a larger composite system that includes the carriers of the information about the past outcomes, while these carriers would be seen as outputs of suitably extended operations in the past. The very notion of operation in the circuit framework, by definition, carries the idea that the input and output systems of an operation are the sole mediator through which any correlations between the outcomes of that operation and the outcomes of other operations is established. One can figuratively think of the boxes in a circuit as isolated space-time regions that can exchange information with each other only through the wires. We will refer to this idea as the `\textit{closed-box}' assumption.

This is not an assumption that concerns the circuit framework as a mathematical model. It is an assumption that we make about a part of an experiment when we say that it corresponds to a valid operation with given input and output systems. One way to think about it is to imagine that we could block the information transmission through the wires corresponding to the assumed input and output systems of the box. If the wires are blocked, all random variables in the box should be completely uncorrelated from other events in the experiment. [Note that a wire in the circuit picture is not a physical wire in space, but is more akin to an ideal channel in time, albeit an instantaneous one (actual channels are represented by boxes, while a wire connecting two boxes signifies that the output of one box is an input of the other).]

If systems were classical objects that we could track as they go from one experiment to another, we could in principle know the paths of information exchange by tracking the systems. But at the microscopic level, the exchange of information is not evident, and in practice a specific circuit structure in the sense above can only be assumed to hold, usually based on some physical considerations (e.g., extrapolating intuition from classical physics). Assuming that we can recognize the structure of boxes and wires in a given experiment, however, is necessary in order for a theory to make non-trivial sense. 

\subsubsection{The no-post-selection criterion in the standard approach}

Consider an experiment in which a quantum operation from system $A$ to system $B$ is chosen out of a set of possible such operations $\{\mathcal{M}_{i_\lambda}^{\lambda, A\rightarrow B}\}_{i_{\lambda}\in O_{\lambda}}$, labeled by $\lambda \in \Lambda$, according to a probability distribution $p(\lambda)\geq 0$, $\sum_{\lambda\in \Lambda} p(\lambda) = 1$, independent of past events. Notice that the whole experiment can be equivalently viewed as a single quantum operation with a larger number of outcomes, $\{p(\lambda)\mathcal{M}_{i_\lambda}^{\lambda, A\rightarrow B}\}_{i_{\lambda}\in O_{\lambda},\lambda\in \Lambda}$. Imagine that the experiment is performed inside a box with input system $A$ and output system $B$ by an apparatus which outputs in separate registers the values of $\lambda$ and $i_\lambda$. If we only look at the value of $\lambda$, conditionally on that information we can say that the operation $\{\mathcal{M}_{i_\lambda}^{\lambda, A\rightarrow B}\}_{i_{\lambda}\in O_{\lambda}}$ has been applied, with the outcome of that operation being stored in the other register. We see that in this case an operation $\{\mathcal{M}_{i_\lambda}^{\lambda, A\rightarrow B}\}_{i_{\lambda}\in O_{\lambda}}$ corresponds to a set of events between $A$ and $B$ that is a \textit{proper subset} of a larger set of possible events that also define an operation. Specifically, the operation  $\{\mathcal{M}_{i_\lambda}^{\lambda, A\rightarrow B}\}_{i_{\lambda}\in O_{\lambda}}$ is obtained from $\{p(\lambda)\mathcal{M}_{i_\lambda}^{\lambda, A\rightarrow B}\}_{i_{\lambda}\in O_{\lambda},\lambda\in \Lambda}$ by updating the description of possible events, conditionally on gained information. This observation suggests a point of view according to which an operation represents knowledge about the possible events in a given region, which can be updated upon learning or discarding of information.

If we have a given operation, any subset of the outcomes of that operation would obey the closed-box assumption, because learning whether the outcome belongs to a subset can be done inside a closed box. However, not all subsets of the outcomes of an operation define a valid operation in the standard formulation of quantum theory. What is the criterion that defines a given set of events that obeys the closed-box assumption as a valid operation?

Intuitively, one may say that the key distinction between a valid operation and an arbitrary subset of the outcomes of a valid operation is that an operation is something that an experimenter is able to choose `at will'. For example, we think that we could `choose' a measurement, but not the outcome of a measurement. How can we formalize this idea?

One possible way could be through the notion of a \textit{freely chosen} variable as a variable that is uncorrelated from events in the past \cite{Bell2,ColbeckRenner, OCB}. If we adopt this point of view, a valid operation would be one that satisfies the causality axiom \cite{CDP2}. But then the axiom would be part of the definition of operation and not an axiom. However, it is clear that there is some nontrivial physical property that this condition expresses, i.e., it is regarded as an axiom for a reason. The fact that theories that allow signaling from the future are considered sensible \cite{noncausal} further shows that the causality axiom is not the implicit assumption that one makes about what an operation is supposed to mean. 

There is another idea that can be said to express the intuition of being able to `choose' an operation, which is in principle compatible with signaling to the past, and in the context of which the causality axiom expresses a nontrivial physical constraint. This is the idea that the variables that define an operation are (or at least can be) obtained \textit{without post-selection}, i.e., they can be known before the time of the input system unconditionally on information available in the future. It is this criterion that is implicitly assumed in the standard approach to quantum theory. Indeed, under this condition the causality axiom captures a nontrivial empirical fact---namely, that out of the larger class of potential operations that we could obtain by selecting sets of events that obey the closed-box assumption, only the class of standard operations can be obtained without post-selection. \textit{A priori}, the non-standard operations that in practice can only be obtained with post-selection could have been obtainable without post-selection too. The fact that they are not can be thought of as enforced by the axiom, since adding any of these operations to the class of standard operations would violate the axiom (the joint probabilities for circuits of such operations can be easily calculated using Bayes's theorem; see Methods).

We remark that if we redefine the class of valid operations not only by adding non-standard operations that require post-selection to the standard ones, but also by excluding standard operations, we can obtain a new class that obeys the causality axiom. For example, we could define measurements to consist only of sets of standard effects $\{E^A_j\}_{j\in Q}$  that satisfy $\sum_{j\in Q} E^A_j=\overline{E}^A$ for some fixed $\overline{E}^A<\id^A$). But the new operations will be isomorphic to operations in the standard theory. In this example, we could redefine the operators describing a measurement as $E^A_j\rightarrow (\overline{E}^A)^{-\frac{1}{2}} E^A_j (\overline{E}^A)^{-\frac{1}{2}} $ and the operators describing a preparation $\{ \rho^A_i \}_{i\in O} $ as  $\rho^A_i \rightarrow    \frac{(\overline{E}^A)^{\frac{1}{2}} \rho^A_i   (\overline{E}^A)^{\frac{1}{2}}}{\tr[\overline{E}^A\overline{\rho}^A ]} $, where $\overline{\rho}^A = \sum_{i\in O}\rho^A_i$, which would put the new preparations and measurements in the standard form while preserving their joint probabilities.

\subsection{The case of classical operational probabilistic theory}

The case of classical OPT in the time-symmetric approach can be obtained by restricting the quantum formalism to CP maps that are diagonal in a fixed orthonormal basis for each system. The general form of such CP maps is given by Eqs.~(30) and (31) in Methods. Here, we discuss the classical case in the context of a simple example, highlighting an important difference between the concept of probability distribution of a classical random variable and the state of a classical system that carries the variable, which may sometimes lead to confusion.

Consider a random bit $X = \{0,1\}$ whose possible values have probabilities $p(X=0)$ and $p(X=1)$, $p(X=0)+p(X=1)=1$, conditionally on some information. According to the circuit framework, well defined probabilities are associated only with the outcomes of closed circuits, so such a random bit must be associated with the two outcomes of an operation from the trivial system to itself. Nevertheless, such a bit must be carried by some physical system, and we may ask what the state of that system can be. We will illustrate the fact that there is no unique answer, as this depends on the form of the measurement by which the value of the bit would be learned. 

First, consider how the logical values $0$ and $1$ can be encoded in a physical system. Since it must be possible to distinguish these values by a suitable measurement on the system, the system has to be non-trivial, i.e., at least two-dimensional. For simplicity, we will assume that it is exactly two-dimensional (we can always think of the relevant information as contained in a suitably defined two-dimensional subsystem of a fictitious larger system). Representing for convenience the states of the system using operators over a two-dimensional Hilbert space that are diagonal in the basis $\{|0\rangle, |1\rangle\}$, one can see that the two logical values must be associated with the two deterministic states $S_0=(|0\rangle \langle 0|; |0\rangle \langle 0|) $ and $S_1=(|1\rangle\langle 1|; |1\rangle \langle 1|)$. This is because only these deterministic states are perfectly distinguishable, i.e., only for them does there exist a measurement $\{E_j\}_{j=0}^1$, such that the probability for its outcomes conditional on the preparation that produces the state (Eq. (21) in Methods) satisfies $p(j|\{E_j\}_{j=0}^1, S_i) = \delta_{i,j}$, $\forall i, j = 0,1$. (Note that here we allow non-standard measurements too.) The most general form of a measurement that achieves this is 
\begin{gather}
\{E_j\}_{j=0}^1, \hspace{0.2cm} \textrm{where} \hspace{0.2cm} E_0 = q |0\rangle\langle 0|, \hspace{0.2cm} E_1 = (2-q) |1\rangle\langle 1|, \hspace{0.2cm} q\in (0,2). \label{measdist}
\end{gather}

Consider now the case in which the value of the bit is not known. We can think that someone has prepared the system that carries the bit in one of the two possible deterministic states $S_0$ or $S_1$, but we do not know which one. This means that the system that carries the bit can be thought of as the output system of a preparation device with two possible outcomes, $i=0,1$, which are such that conditionally on knowledge of the outcome, the preparation would be updated to the respective deterministic preparation $\{|0\rangle \langle 0|\}$ or $\{|1\rangle\langle 1|\}$ (corresponding to the deterministic state $S_0$ or $S_1$). From the rule for updating the description of an operation conditionally on information about its outcome (Eq. (34) in Methods and discussion after it), we see that such a preparation can most generally have the form $\{\rho_i\}_{i=0}^1$, where $\rho_0= p|0\rangle\langle 0|$, $\rho_1 = (1-p)|1\rangle\langle 1|$, $p\in [0,1]$, with the two states corresponding to the two outcomes being respectively $(\rho_0; \overline{\rho})$ and $(\rho_1; \overline{\rho})$, $\overline{{\rho}}= p|0\rangle\langle 0|+(1-p)|1\rangle\langle 1|$. If we have no access to the outcomes of the preparation, we would describe it by the coarse-grained deterministic preparation $\{\overline{\rho}\}$ yielding the deterministic state $(\overline{\rho}; \overline{\rho})$.

In the standard approach, where all measurements are assumed to be of the standard type and the theory obeys the causality axiom, the coefficients $p$ and $1-p$ above are equal to the probabilities of the respective preparation outcomes, for any measurement applied on the system. They are also equal to the probabilities of the outcomes of a subsequent measurement on the system that reads out the value of the bit (there is only one such measurement in the standard approach: $E_0= |0\rangle\langle 0|$, $E_1= |1\rangle\langle 1|$). This is why the standard density operator $\overline{\rho} = p|0\rangle\langle 0|+(1-p)|1\rangle\langle 1|$ is often regarded as equivalent to our probabilistic description of the random bit (where $p(X=0)=p$ and $p(X=1)=1-p$). However, we stress that in the general case the coefficients $p$ and $(1-p)$ are \textit{not} equal to the probabilities of the outcomes of the preparation, nor to the probabilities of a subsequent measurement on the system that reads the value of the bit. Indeed, consider a measurement of the form \eqref{measdist} with $q\neq 1$ (obtained, for instance, using post-selection as described below). In this case, the joint probabilities for the preparation and measurement outcomes are 
\begin{gather}\label{joint}
p(i,j|\{\rho_i\}_{i=0}^1, \{E_j\}_{j=0}^1) = 
\begin{cases}
\frac{pq}{pq+(1-p)(2-q)}, \hspace{0.2cm} \textrm{for} \hspace{0.2cm} i=j=0\\
\frac{(1-p)(2-q)}{pq+(1-p)(2-q)}, \hspace{0.2cm} \textrm{for} \hspace{0.2cm} i=j=1\\
0, \hspace{1.7cm} \textrm{for} \hspace{0.2cm} i\neq j
\end{cases}.
\end{gather}
Note that, since the measurement is one that perfectly distinguishes the two values of the bit, the (marginal) probability of the event of preparing a given value $x$ of the bit $X$ is equal to the probability of the event of reading out that value, and equal to the probability of the joint event of preparing and reading out that value. The logical value of the bit $X$ can thus be associated with any of these respective events, and its probability with the probability of that corresponding event. We have
\begin{gather}
p(X=0)= \frac{pq}{pq+(1-p)(2-q)}, \notag\\
p(X=1) = \frac{(1-p)(2-q)}{pq+(1-p)(2-q)}.\label{probprep}
\end{gather} 
We see that there is no unique value of $p$ (and hence no unique deterministic state $(\overline{\rho}; \overline{\rho})$) associated with a system carrying a random bit with a given probability distribution $p(X)$, unless we also specify the form of the measurement (here the parameter $q$), which in the standard approach is implicitly specified (the case $q=1$). 

The difference between the state of a classical system and the probability of the random variable carried by the system can be seen clearly in the context of updating the probabilities of the outcomes of the preparation conditionally on information about the outcomes of the measurement. Consider the same preparation as above, but connected to the standard three-outcome measurement
$\{E'_k\}_{k=0}^2$, where $E'_0 = \frac{q}{2}|0\rangle\langle 0|$, $E'_1=\frac{2-q}{2}|1\rangle\langle 1|$, $E'_2=\frac{2-q}{2}|0\rangle\langle 0|+ \frac{q}{2}|1\rangle\langle 1|$. In this case, the probabilities of the two preparation outcomes are $p$ and $(1-p)$. (The joint probabilities of the preparation and measurement outcomes are  $p(i=0,k=0|\{\rho_i\}_{i=0}^1, \{E'_k\}_{k=0}^2) = {pq}/{2}$, $p(i=0,k=1|\{\rho_i\}_{i=0}^1, \{E'_k\}_{k=0}^2) = 0$, $p(i=0,k=2|\{\rho_i\}_{i=0}^1, \{E'_k\}_{k=0}^2) = {p(2-q)}/{2}$, $p(i=1,k=0|\{\rho_i\}_{i=0}^1, \{E'_k\}_{k=0}^2) = 0$, $p(i=1,k=1|\{\rho_i\}_{i=0}^1, \{E'_k\}_{k=0}^2) = {(1-p)(2-q)}/{2}$, $p(i=1,k=2|\{\rho_i\}_{i=0}^1, \{E'_k\}_{k=0}^2) = {(1-p)q}/{2}$.) If we look at the outcome of the measurement in a way that only reveals whether the outcome is $k=2$ or not, conditionally on finding out that it is not, we would update the description of the measurement exactly to the two-outcome measurement \eqref{measdist}. The joint probabilities for the preparation and measurement outcomes are also updated accordingly, in agreement with Bayes's theorem, to those in Eq.~\eqref{joint}, with the probabilities of the preparation outcomes becoming those in Eq.~\eqref{probprep}. Note, however, that even though the probabilities of the preparation outcomes are updated, the state associated with the coarse-grained preparation is not. This is because the state, by definition, is a mathematical object associated with the local description of the procedure in the preparation box, and this description is not altered by information gained from the measurement box. Since in the standard approach the state and the probability of the preparation outcomes are often identified, it is common to see discussions about updating the state that the system had prior to a measurement conditionally on the outcome of the measurement. From the perspective of the operational approach, this is a category mistake. The state of a system can only be updated conditionally on information gained from the preparation box.

\subsection{Consistency of the operational interpretation of the time-symmetric formulation}

As discussed in the main text, in the limit where all physical systems in the universe are included in our description, we obtain a global `field' picture similar to the one in Fig. 4, where the transformation in the bulk of the space-time region between two instants of time is deterministic, and all random data is outsourced to the boundary. The toy example of Fig. 4 depicts the case where the dynamics is unitary and the future boundary measurement is of the standard type, but the time-symmetric formulation in principle permits more general operations both in the bulk and on the boundary, because it does not require the sum of the CP maps associated with the outcomes of an operation to be a CPTP map. For example, Fig. 5 illustrates the case of unitary dynamics in the bulk combined with a non-standard future boundary measurement, and how this leads to the possibility of effectively obtaining non-standard operations at specific space-time location without post-selection. One may wonder whether such scenarios make sense operationally because, for instance, the possibility for a non-standard future boundary condition allows for non-local correlations to be established as a result of events in the future, and this seems to offer the possibility of explaining arbitrary observations as a result of suitable future boundary conditions, rendering the theory non-falsifiable. Here, we discuss why this is not the case, highlighting the fact that the theory makes operational sense locally while also being consistent with the global field picture.  

The key point to be emphasized is that in practice we infer the global picture based on the results of local experiments described by locally available information, not the other way around. As noted earlier, the assumption that we can recognize experimental setups corresponding to specific circuits is necessary in order for an operational probabilistic theory to have an empirical meaning. By definition, a circuit is associated with an experimental setup in which specific events have well defined probabilities conditionally on the variables that define the setup only. Hence, a local circuit makes sense by definition. The local circuits that we find in practice do not have to be of the standard kind---we may find circuits consisting of non-standard operations that can be obtained without post-selection (as in the situation depicted in Fig. 5). But any local circuit should be consistent with the global field picture, i.e., it should be possible to understand it as arising effectively from the global circuit of the universe, even though we do not need to know the global circuit in order to describe the local experimental setup and corresponding local circuit. Note that according to the global field picture all classical information in the universe can be thought of as existing on a holographic hypersurface (the boundary of space-time), but consistently with it, we can also think of effective local circuits taking place in the bulk. By definition, any classical information that can be thought to exist in the bulk must be projected consistently on the boundary. In particular, the probabilities for all classical variables that can be thought to exist in the bulk are the same as those of their holographic projections. Thus, the local and global points of view are consistent. 

Finally, we remark that similarly to the standard formulation of quantum theory, the time-symmetric formulation makes falsifiable propositions. As pointed out, e.g., in Ref.~\cite{Perinotti}, falsifiable propositions are introduced in the theory by the existence of states that can be perfectly distinguished from some other states. The fact that the time-symmetric formulation contains such logical propositions was demonstrated for the special case of diagonal operators in the previous section. More generally, any deterministic state $(\overline{\rho}^A; \overline{\rho}^A)$, where $\overline{\rho}^A\in \mathcal{L}(\mathcal{H}^A)$ does not have full rank, can be perfectly distinguished from a state $(\overline{\sigma}^A; \overline{\sigma}^A)$, where $\overline{\sigma}^A$ has support orthogonal to the support of $\overline{\rho}^A$. We can distinguish two such states using, for example, a measurement of the standard type with two outcomes corresponding to the effects $(P^A; \id^A)$ and $(\id^A-P^A; \id^A)$, where $P^A$ is the projector on the support of $\overline{\rho}^A$. \\

\textbf{Acknowledgements.} This work was supported by the European Commission under the Marie Curie Intra-European Fellowship Programme (PIEF-GA-2010-273119) and by the F.R.S.--FNRS under the Charg\'{e} de recherches (CR) Fellowship Programme.

\end{document}